\documentclass[12pt]{elsart}

\usepackage{graphicx}
\usepackage{subfigure}
\usepackage{amsmath}
\usepackage{amssymb}
\usepackage{cite}

\begin{document}

\begin{frontmatter}

\title{Phase boundaries of a spin-3/2 Blume-Emery-Griffiths model on a honeycomb lattice}
\author{M. \v{Z}ukovi\v{c}},
\ead{milan.zukovic@upjs.sk}
\author{M. Ja\v{s}\v{c}ur}
\address{Department of Theoretical Physics and Astrophysics, Faculty of Science,\\ 
P. J. \v{S}af\'arik University, Park Angelinum 9, 041 54 Ko\v{s}ice, Slovakia}

\begin{abstract}
The spin-3/2 Blume-Emery-Griffiths model on a honeycomb lattice is studied by Monte Carlo simulations with the goal to determine phase diagrams for a range of the model parameters and to investigate the nature of the phase transitions between the respective phases. For positive values of the biquadratic to bilinear interaction ratio $\alpha$, we find two ferromagnetically ordered phases, $F_1$ and $F_2$, with the sublattice magnetizations $(1/2,1/2)$ and $(3/2,3/2)$, respectively, and our results confirm the discontinuous character of the order-disorder critical line as a function of the single-ion anisotropy strength, predicted by the effective-field theory (EFT). For negative values of $\alpha$, there is another ferrimagnetic ($FRM$) phase of the type $(1/2,3/2)$, located between $F_1$ and $F_2$. However, the step-like variation of the order-disorder critical frontier obtained from EFT for large negative $\alpha$ is not reproduced and thus deemed artifact of the EFT approximation. Finite-size scaling analysis performed at various points between the respective identified phases gave the ratio of critical exponents $\gamma/\nu$ consistent with the 2D Ising universality class, except in the vicinity of the boundary intersection, where the results deviated from the standard values beyond the measurement errors.
\end{abstract}


\end{frontmatter}

\section{Introduction}
\hspace*{5mm} The spin-$S$ Blume-Emery-Griffiths (BEG) model is a spin-$S$ Ising model which besides bilinear exchange interactions also includes biquadratic interactions and a single-ion anisotropy. It was introduced to understand behavior of some real physical systems, such as helium mixtures~\cite{blum71} and metamagnets ($S=1$) or ternary mixtures and compound $\rm{DyVO}_4$ ($S=3/2$). The spin-3/2 BEG model was studied by various approaches, including the mean field theory (MFT)~\cite{siva72,krin75,saba91,plas93}, the effective-field theory (EFT)~\cite{kane93,bakk96,peli96}, the renormalization group (RG)~\cite{bakc93,bakc01}, the two-spin cluster expansion (CE)~\cite{ilko96}, the cluster variation method (CVM)~\cite{tuck00,kesk08}, the pair approximation (PA), Monte Carlo (MC) simulations~\cite{lara98} and cellular automata (CA)~\cite{sefe10}. \\
\hspace*{5mm} Despite intensive investigations, the critical behavior of the model is still not well understood. Even in the most studied case with zero biquadratic interactions, i.e., the Blume-Capel (BC) model~\cite{blum66,cape66,cape67}, no consensus among various approaches has been established. In particular, the MFT results, supported by some preliminary Monte Carlo (MC) simulations~\cite{saba91}, as well as the EFT~\cite{peli96} and RG~\cite{bakc93} calculations, identified at low temperatures the presence of two ferromagnetic phases ${\rm F_1}$ and ${\rm F_2}$ with the ground-state sublattice magnetization structures $(1/2,1/2)$ and $(3/2,3/2)$, respectively. The two phases were claimed to be separated by a first-order phase boundary that extends all the way to the second-order line which forms a phase boundary with the paramagnetic phase at higher temperatures. This scenario was confronted by conclusions from later MFT~\cite{plas93}, two-spin cluster expansion~\cite{ilko96}, MC~\cite{lara98,bekh97}, and CVT~\cite{tuck00} investigations, which predicted that the first-order line at low temperatures did not extend up to the paramagnetic phase boundary line, but terminated at an isolated point. \\
\hspace*{5mm} The spin-3/2 BEG model with finite biquadratic interactions was much less investigated. Nevertheless, besides the two ferromagnetic phases ${\rm F_1}$ and ${\rm F_2}$ observed in the BC model, for a certain range of the biquadratic interactions and the singe-ion anisotropy strength it is expected to display a ferrimagnetic (${\rm FRM}$) phase with unequal sublattice magnetizations in the ground state, i.e., $(1/2,3/2)$~\cite{saba91,bakk96}. However, it is not clear what is the nature of the phase transitions from the paramagnetic to the ferrimagnetic phase and between the respective ordered phases. Another mysterious feature is a step-wise variation of the critical temperature for larger negative values of the biquadratic interactions obtained within the EFT approach on a honeycomb lattice~\cite{kane93}. To our best knowledge, this peculiarity has not been explained neither verified by any other alternative approach and, therefore, it is not known whether it is real or just an artifact of the used approximation. Furthermore, for the present model with finite biquadratic interactions we can also define a ferriquadrupolar order parameter and study phase transitions to the  ferriquadrupolar (${\rm FRQ}$) phase. The MFA~\cite{siva72} and CA~\cite{sefe10} studies predicted the existence of separate magnetic and ferriquadrupolar phase transitions, leading to rich phase diagrams. \\
\hspace*{5mm} Motivated by the above findings, we perform Monte Carlo simulations of the spin-3/2 BEG model on a honeycomb lattice in order to determine phase diagrams for a range of the model parameters and to investigate the nature of the phase transitions between the respective phases. 

\section{Model and methods} 
\hspace*{5mm} The spin-3/2 Blume-Emery-Griffiths model on a honeycomb lattice can be described by the Hamiltonian 
\begin{equation}
\label{Hamiltonian}
H=-J_1\sum_{\langle i,j \rangle}S_{i}S_{j}-J_2\sum_{\langle i,j \rangle}S_{i}^2S_{j}^2-D\sum_{i}S_{i}^2,
\end{equation}
where $S_{i}=\pm 3/2, \pm 1/2$ is a spin on the $i$th lattice site, $\langle i,j \rangle$ denotes the sum over nearest neighbors, $J_1>0$ is a ferromagnetic bilinear exchange interaction parameter, $J_2$ is a biquadratic exchange interaction parameter and $D$ is a single-ion anisotropy parameter.

\subsection{Ground state determination}
The honeycomb lattice system is considered to consist of two interpenetrating sublattices A and B. Then, assuming sublattice uniformity we can focus on an elementary unit cell comprising the central spin, let say from the sublattice A, i.e., $S_{A}$, and its three nearest neighbors from the sublattice B, i.e., $S_{B}$, and express its reduced ground-state (GS) energy per spin as 
\begin{equation}
\label{Energy}
e=-\frac{3}{2}S_{A}S_{B}-\frac{3}{2}\alpha S_{A}^2S_{B}^2-\frac{\Delta}{2}(S_{A}^2+S_{B}^2),
\end{equation}
where $\alpha=J_2/J_1$ and $\Delta=D/J_1$.  Then we can distinguish the following states:
\begin{itemize}
	\item  $F_1$ - ferromagnetic state with $S_A=S_B=\pm\frac{1}{2}$ and the energy $e_1=-\frac{3}{8}-\frac{3}{32}\alpha-\frac{\Delta}{4}$,
	\item  $F_2$ - ferromagnetic state with $S_A=S_B=\pm\frac{3}{2}$ and $e_2=-\frac{27}{8}-\frac{243}{32}\alpha-\frac{9\Delta}{4}$, and
	\item  $FRM$ - ferrimagnetic state with $S_A=\pm\frac{1}{2}$, $S_B=\pm\frac{3}{2}$ or $S_A=\pm\frac{3}{2}$, $S_A=\pm\frac{1}{2}$ and $e_3=-\frac{9}{8}-\frac{27}{32}\alpha-\frac{5\Delta}{4}$.
\end{itemize}
GS in different regions of the parameter space $(\alpha-\Delta)$ can be determined from the condition of the minimum energy given by Eq.~(\ref{Energy}).

\subsection{Monte Carlo simulation}
In order to study behavior of various thermodynamic quantities in the parameter space and to determine the phase diagrams, we employ Monte Carlo (MC) method with the Metropolis dynamics and the periodic boundary conditions. For thermal averaging we consider $N=L \times 10^4$ MCS (Monte Carlo sweeps or steps per spin), where $L=24-96$ is the linear lattice size, after discarding additional $20\%$ of MCS for thermalization. To obtain dependencies on the reduced temperature $t \equiv k_BT/J_1$ at a fixed value of $\Delta$, the simulations start from the paramagnetic phase using random initial configurations. Then the temperature is gradually lowered and the new simulation starts from the final configuration obtained at the previous temperature. To obtain variations of the quantities as functions of $\Delta$, we run simulations at a fixed temperature which may start from other than the paramagnetic phase. Thus an appropriate initial state should be chosen, such as all spins in the state 1/2 (3/2) if we start from ${\rm F_1} ({\rm F_2}$) phase. Such an approach ensures that the system is maintained close to the equilibrium in the entire range of the changing parameter and considerably shortens thermalization periods. For reliable estimation of statistical errors, we used the $\Gamma$-method~\cite{wolf04}, which focuses on the explicit determination of the relevant autocorrelation functions and times. It has been shown to produce more certain error estimates than the binning techniques, which handle autocorrelations only implicitly. We note that the $\Gamma$-method allows assessing statistical errors for arbitrary in general nonlinear functions of elementary observables in MC simulations. In order to obtain critical exponents, we perform finite-size scaling (FSS) analysis, using the linear sizes $L=24,48,72$ and $96$, up to $N=10^7$ MCS and employing the reweighting techniques~\cite{ferr88}.\\
\hspace*{5mm} On the honeycomb lattice we calculate respective sublattice dipolar and quadrupolar order parameters per site $m_{\rm X}$ and $q_{\rm X}$ (X = A or B)
\begin{equation}
\label{sub_mag}
m_{\rm X} = 2\langle M_{\rm X}\rangle/L^2 = 2\Big\langle\sum_{i \in {\rm X}}S_{i}\Big\rangle/L^2,
\end{equation}
\begin{equation}
\label{sub_quad}
q_{\rm X} = 2\langle Q_{\rm X}\rangle/L^2 = 2\Big\langle\sum_{i \in {\rm X}}S_{i}^2\Big\rangle/L^2,
\end{equation}
and lattice order parameters $m_d$, $q_d$ (direct) and $m_s$, $q_s$ (staggered)
\begin{equation}
\label{mag}
m_d = \langle M_d \rangle/L^2 = \Big\langle\sum_{i \in {\rm A}}S_{i}+\sum_{j \in {\rm B}}S_{j}\Big\rangle/L^2,
\end{equation}
\begin{equation}
\label{stag_mag}
m_s = \langle M_s \rangle/L^2 = \Big\langle\Big|\sum_{i \in {\rm A}}S_{i}-\sum_{j \in {\rm B}}S_{j}\Big|\Big\rangle/L^2,
\end{equation}
\begin{equation}
\label{quad}
q_d = \langle Q_d \rangle/L^2 = \Big\langle\sum_{i \in {\rm A}}S_{i}^2+\sum_{j \in {\rm B}}S_{j}^2\Big\rangle/L^2,
\end{equation}
\begin{equation}
\label{stag_quad}
q_s = \langle Q_s \rangle/L^2 = \Big\langle\Big|\sum_{i \in {\rm A}}S_{i}^2-\sum_{j \in {\rm B}}S_{j}^2\Big|\Big\rangle/L^2,
\end{equation}
where $\langle\cdots\rangle$ denotes thermal average. Further, we calculate susceptibilities pertaining to the respective lattice order parameters 
\begin{equation}
\label{chi}\chi_{u}^O = \frac{\langle O_{u}^{2} \rangle - \langle O_{u} \rangle^{2}}{L^2k_{B}T}, 
\end{equation}
where $O=M$ or $Q$ and $u=d$ or $s$, specific heat per site $C$
\begin{equation}
\label{c}C=\frac{\langle H^{2} \rangle - \langle H \rangle^{2}}{L^2k_{B}T^{2}},
\end{equation}
logarithmic derivatives of $\langle O_u \rangle$ and $\langle O_{u}^{2} \rangle$ with respect to $\beta=1/k_{B}T$,
\begin{equation}
\label{D1}D_{u1}^O = \frac{\partial}{\partial \beta}\ln\langle O_u \rangle = \frac{\langle O_u H
\rangle}{\langle O_u \rangle}- \langle H \rangle,
\end{equation}
\begin{equation}
\label{D2}D_{u2}^O = \frac{\partial}{\partial \beta}\ln\langle O_{u}^{2} \rangle = \frac{\langle O_{u}^{2} H
\rangle}{\langle O_{u}^{2} \rangle}- \langle H \rangle.
\end{equation}
For the FSS analysis we use the following scaling relations:
\begin{equation}
\label{scalchi}\chi_{u,max}^O(L) \propto L^{\gamma_u^O/\nu_u^O},
\end{equation}
\begin{equation}
\label{scalD1}D_{u1,max}^O(L) \propto L^{1/\nu_u^O},
\end{equation}
\begin{equation}
\label{scalD2}D_{u2,max}^O(L) \propto L^{1/\nu_u^O},
\end{equation}
\noindent where $\nu_u^O$ and $\gamma_u^O$ are the critical exponents of the correlation length and susceptibility, respectively.

\section{Results}
\begin{figure}[t!]
\centering
\subfigure{\includegraphics[scale=0.36,clip]{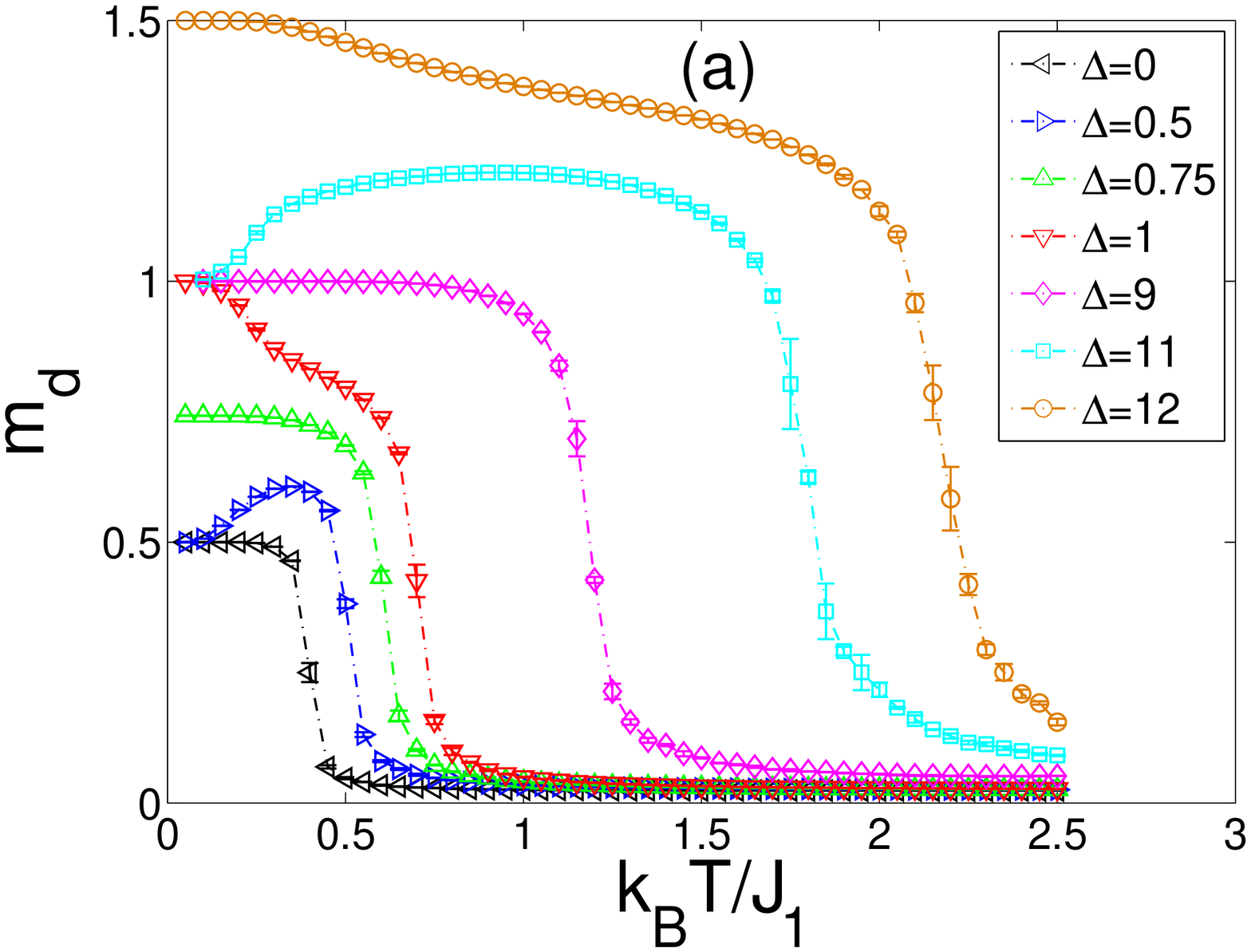}\label{fig:m-T_J-2}}
\subfigure{\includegraphics[scale=0.36,clip]{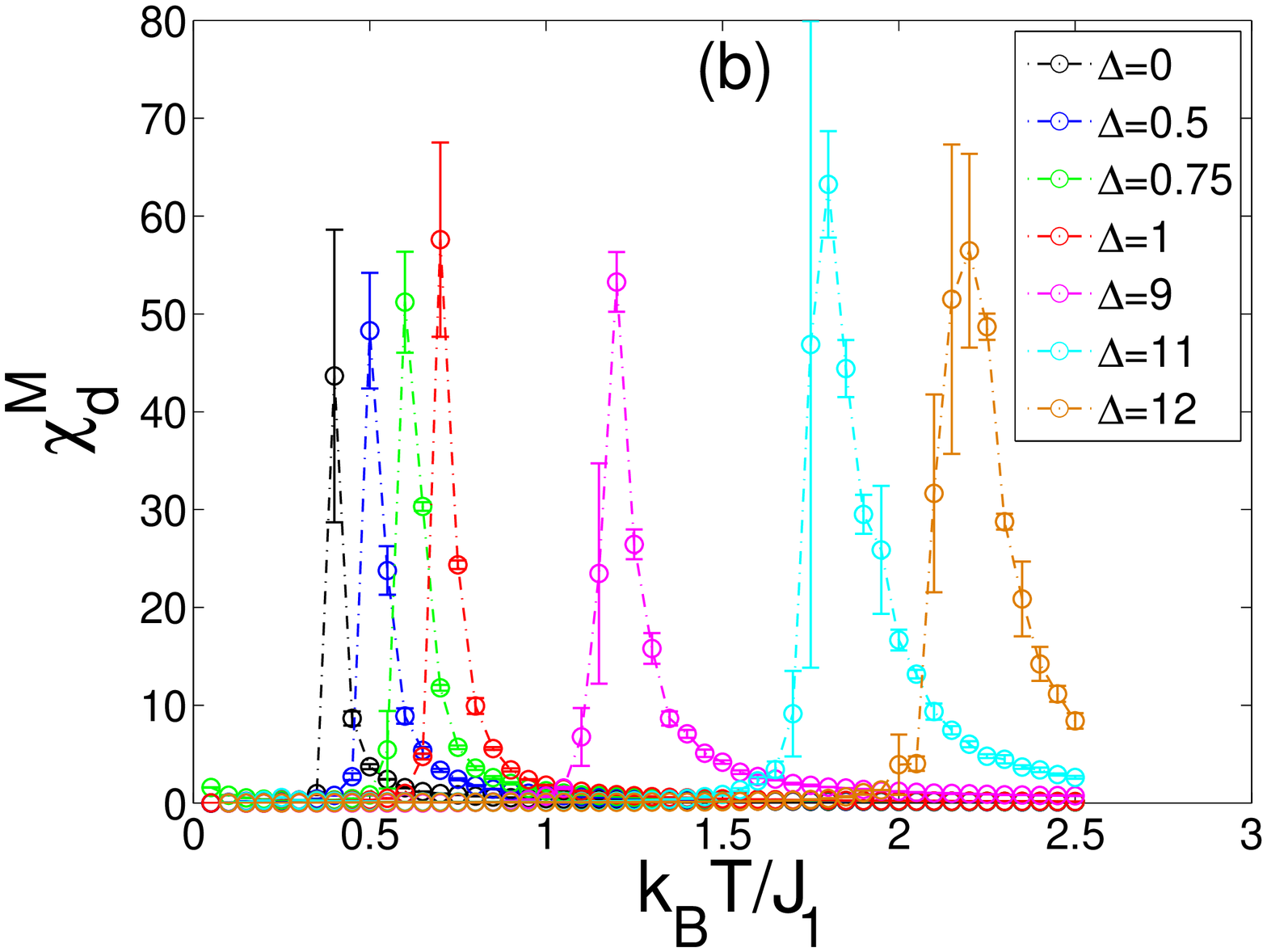}\label{fig:xi-T_J-2}}
\caption{Temperature variation of (a) the direct magnetization $m_d$ and (b) the direct magnetic susceptibility $\chi_d^{M}$, for different values of $\Delta$ and $L=48$.}\label{fig:m_xi-T_J-2}
\end{figure}

\begin{figure}[b!]
\centering
\subfigure{\includegraphics[scale=0.23,clip]{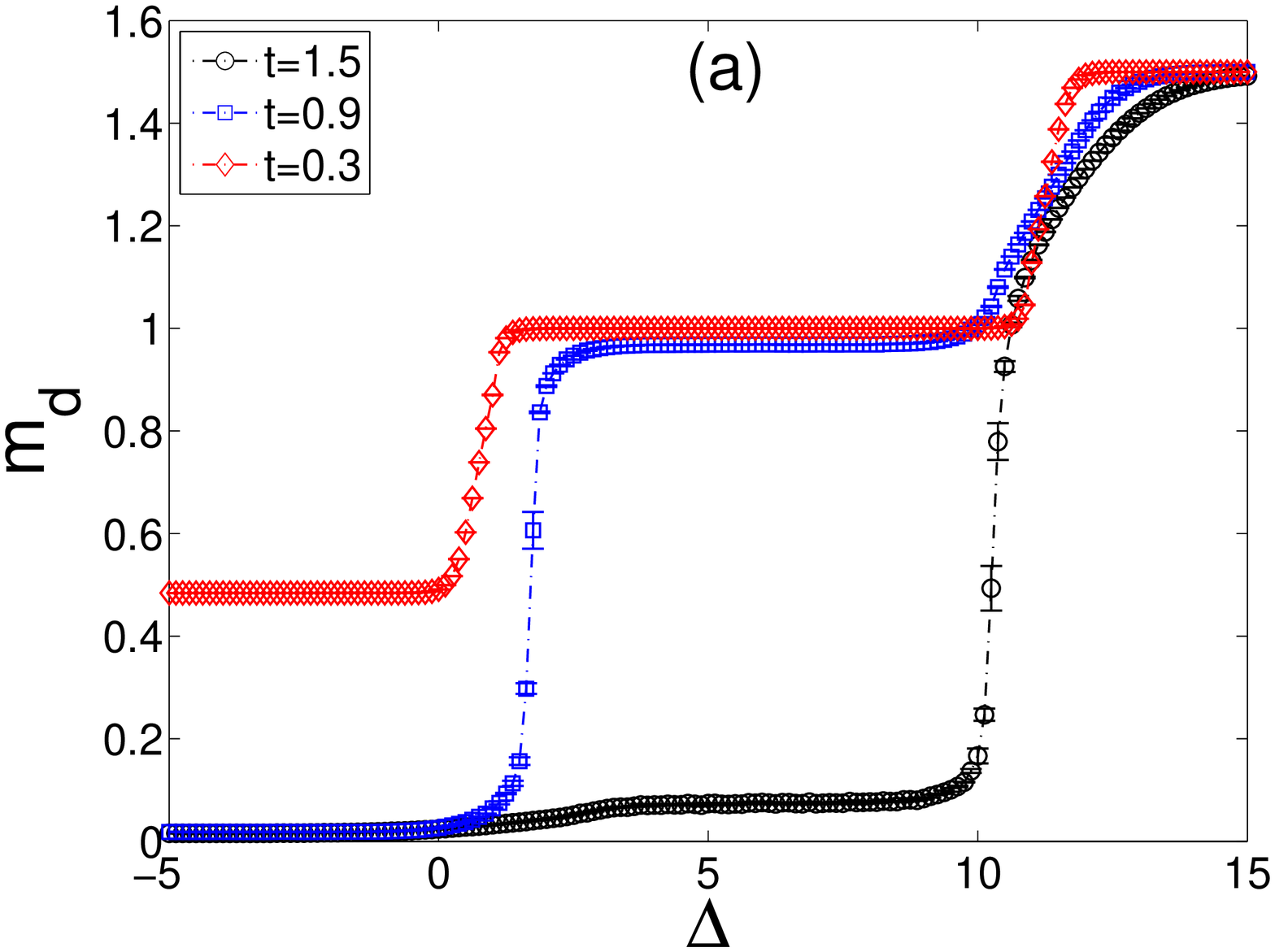}\label{fig:m-D_J-2}}
\subfigure{\includegraphics[scale=0.23,clip]{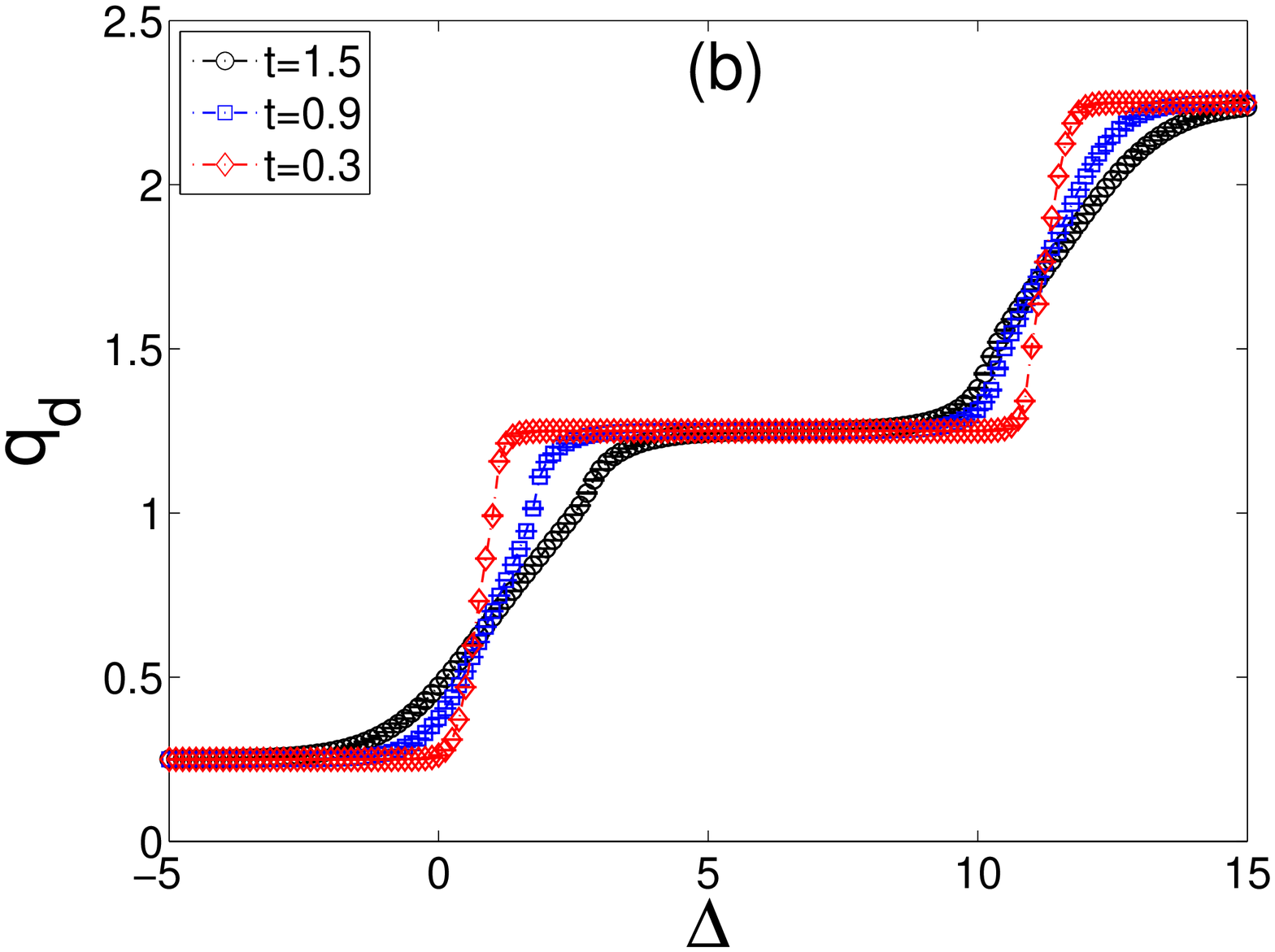}\label{fig:q-D_J-2}}
\subfigure{\includegraphics[scale=0.23,clip]{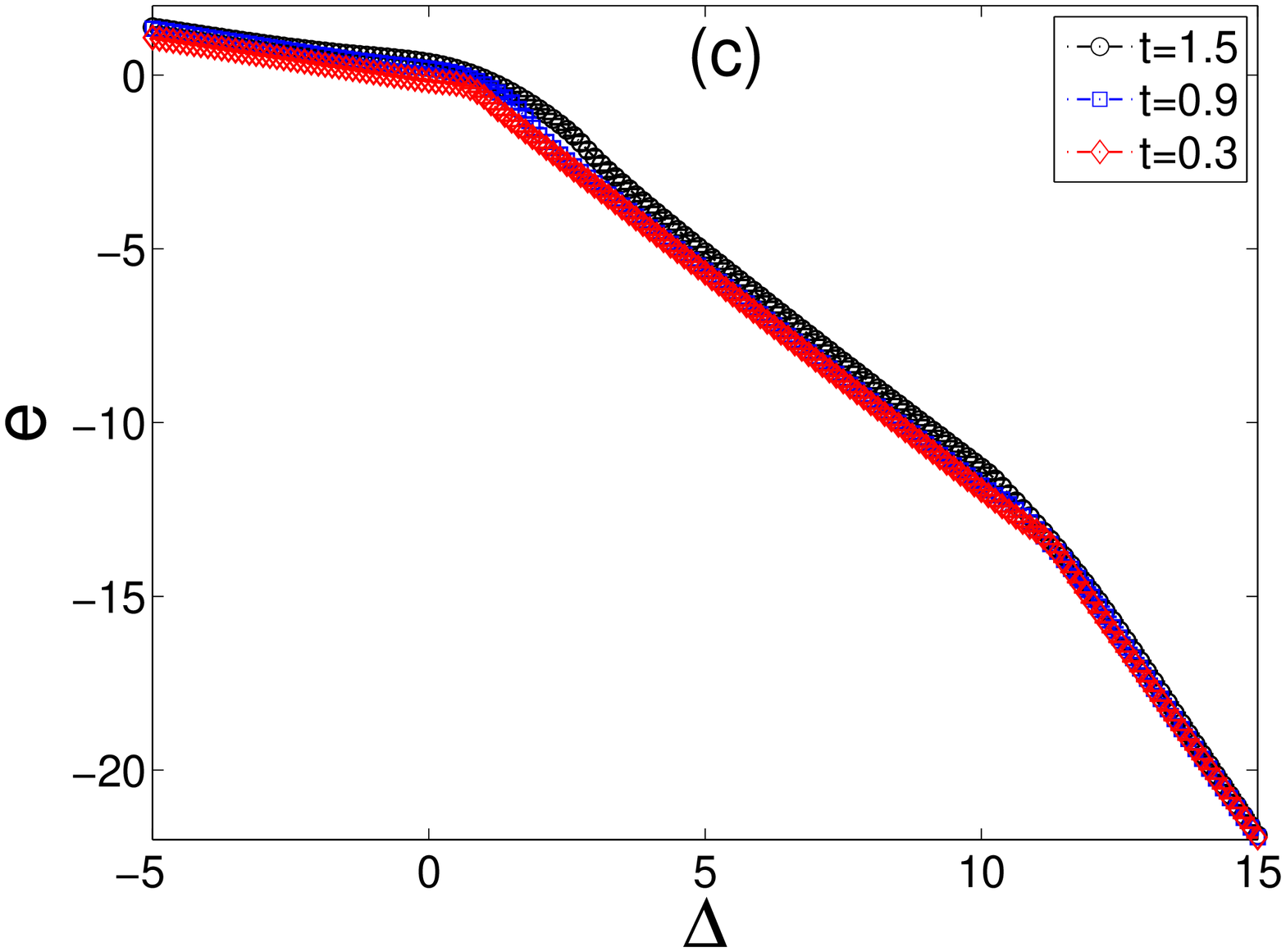}\label{fig:e-D_J-2}} \\
\subfigure{\includegraphics[scale=0.23,clip]{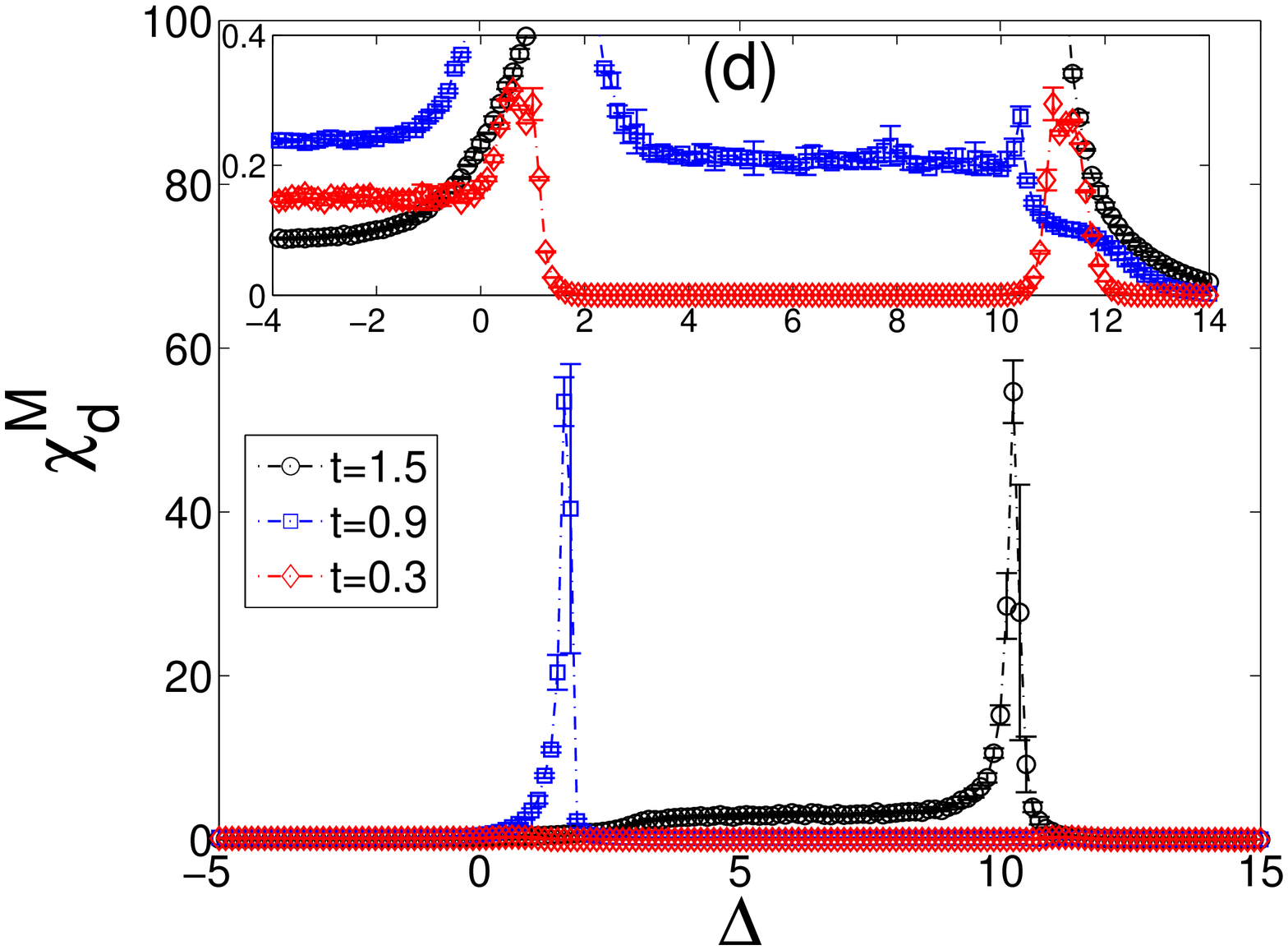}\label{fig:chi-D_J-2}}
\subfigure{\includegraphics[scale=0.23,clip]{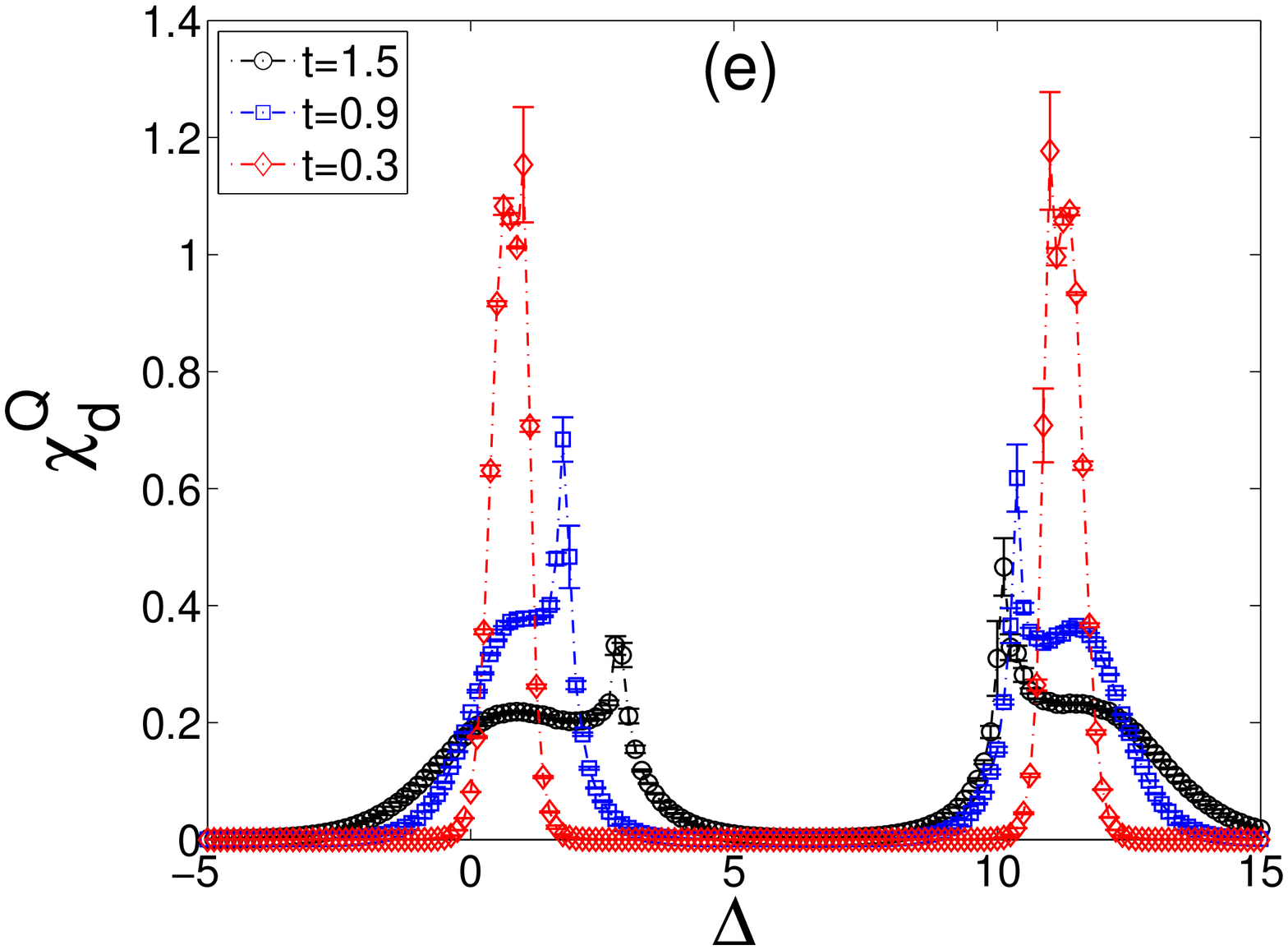}\label{fig:chiq-D_J-2}}
\subfigure{\includegraphics[scale=0.23,clip]{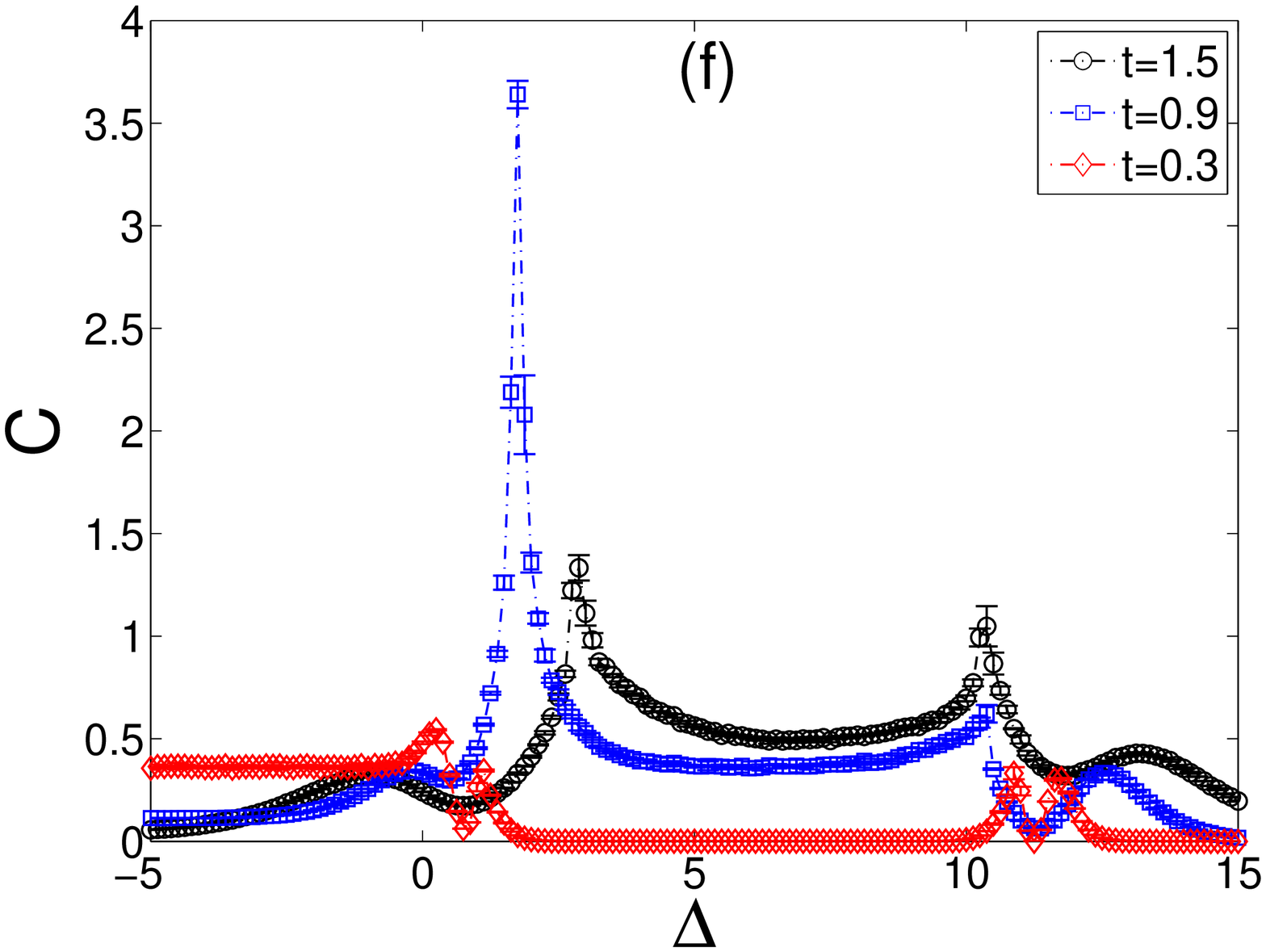}\label{fig:c-D_J-2}}
\caption{Variations of $m_d,q_d$ and $e$ and the corresponding response functions $\chi_d^M, \chi_d^Q$ and $C$, respectively, with the single-ion anisotropy parameter $\Delta$, for different values of $t$ and $L=48$.}\label{fig:x-D_J-2}
\end{figure}

\begin{figure}[t!]
\centering
\includegraphics[scale=0.5,clip]{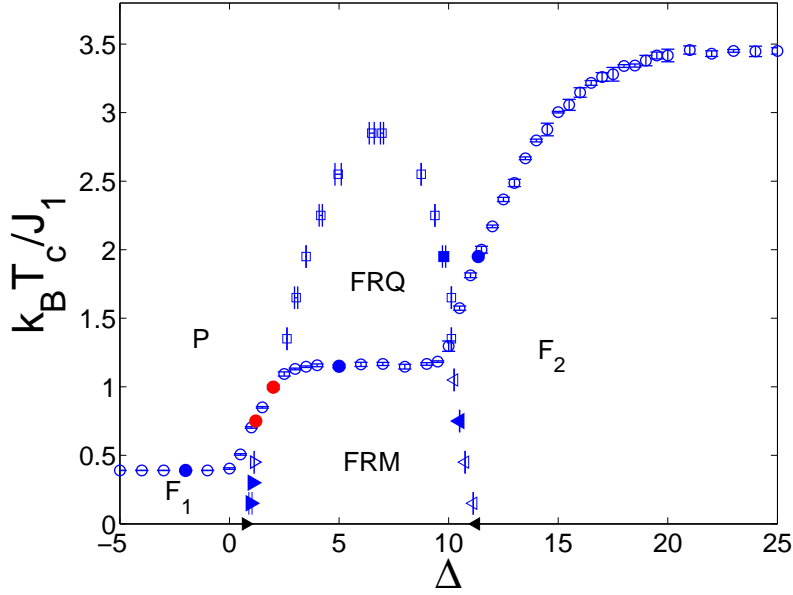}
\caption{Phase diagram in $(t-\Delta)$ parameter space. The filled symbols represent the FSS points and those in red color nonuniversal behavior.}\label{fig:PD_J2-2}
\end{figure}

\begin{figure}[t!]
\centering
\subfigure{\includegraphics[scale=0.25,clip]{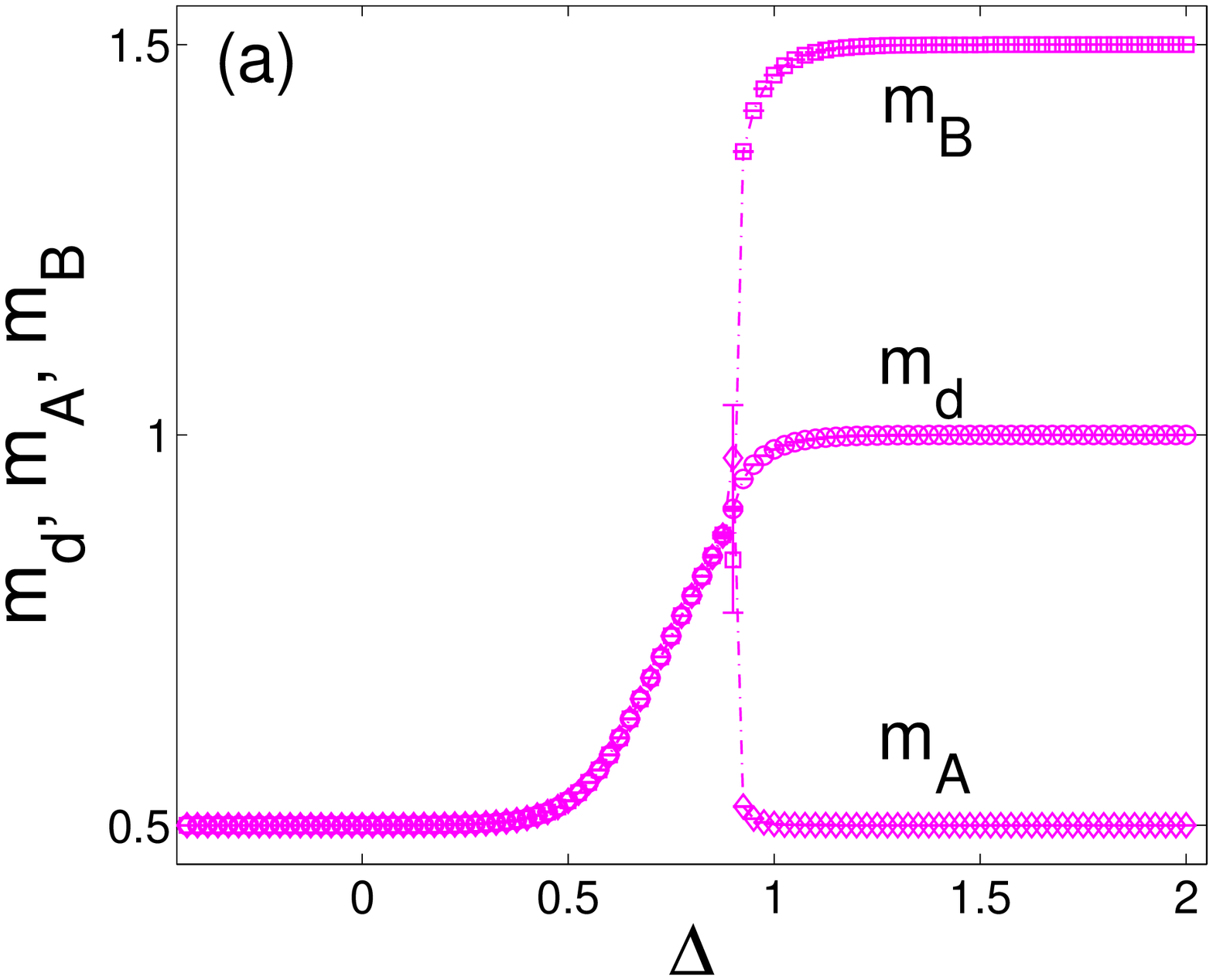}\label{fig:m-D_J-2_t015}}
\subfigure{\includegraphics[scale=0.25,clip]{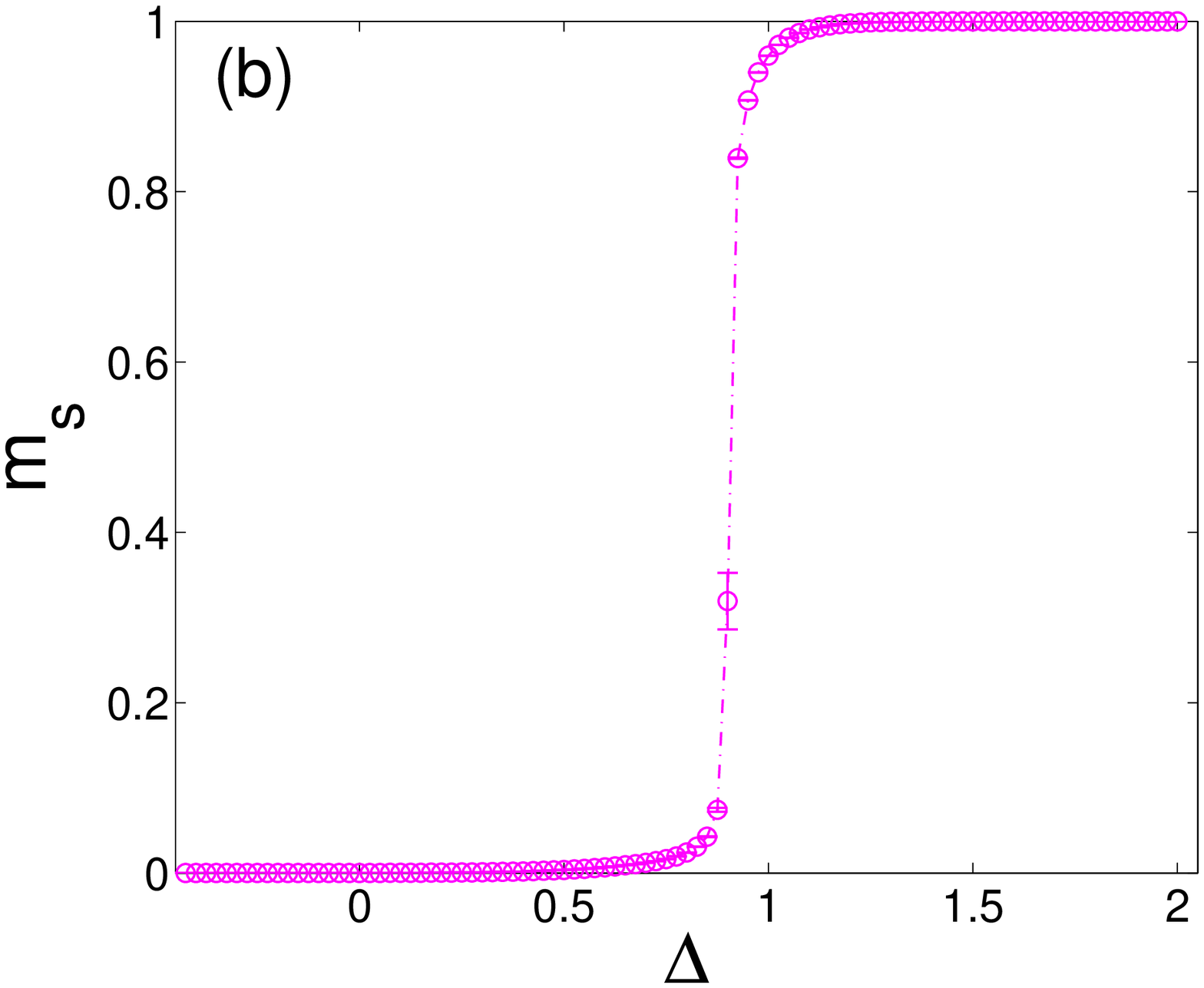}\label{fig:ms-D_J-2_t015}}
\subfigure{\includegraphics[scale=0.25,clip]{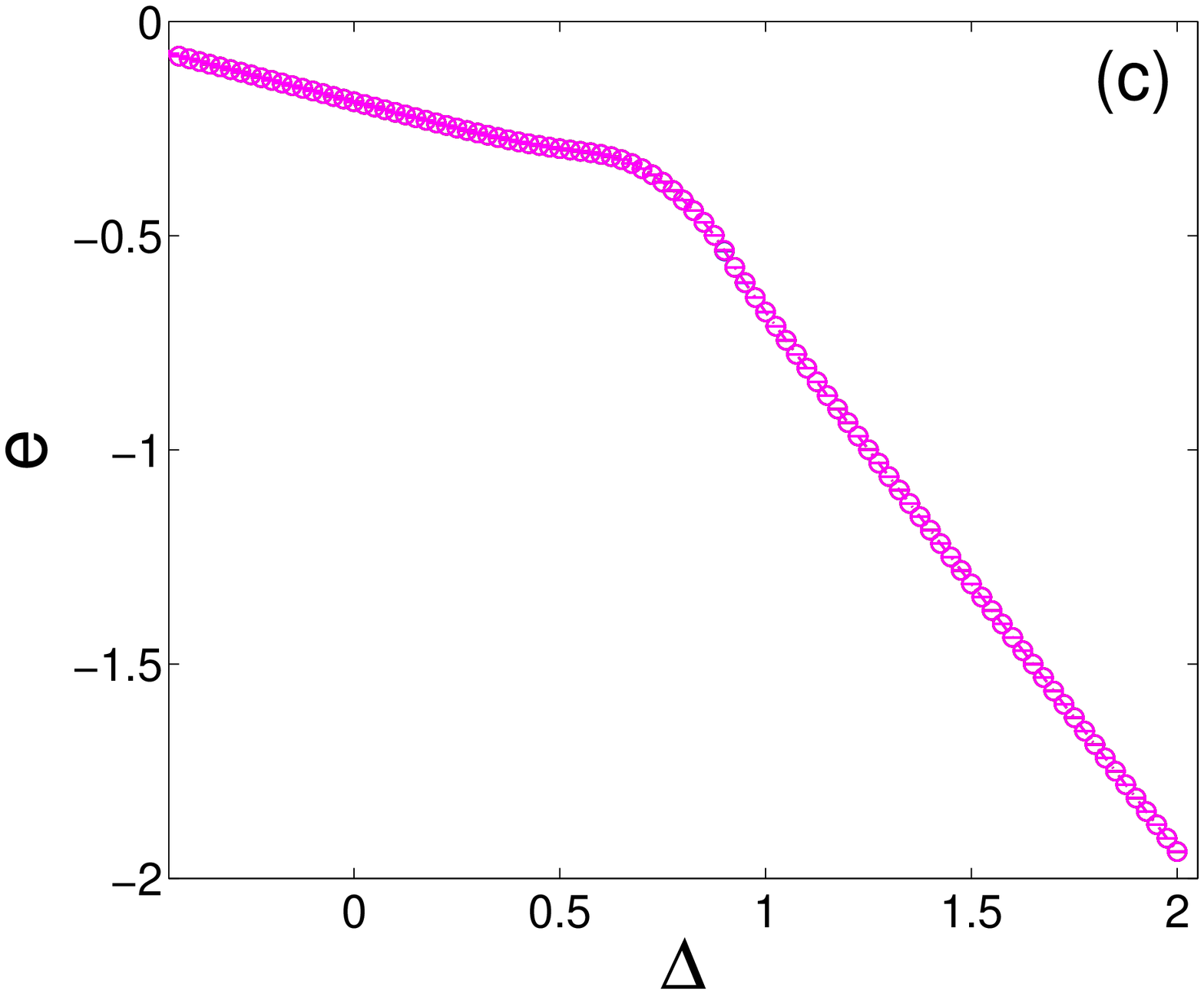}\label{fig:e-D_J-2_t015}} \\
\subfigure{\includegraphics[scale=0.25,clip]{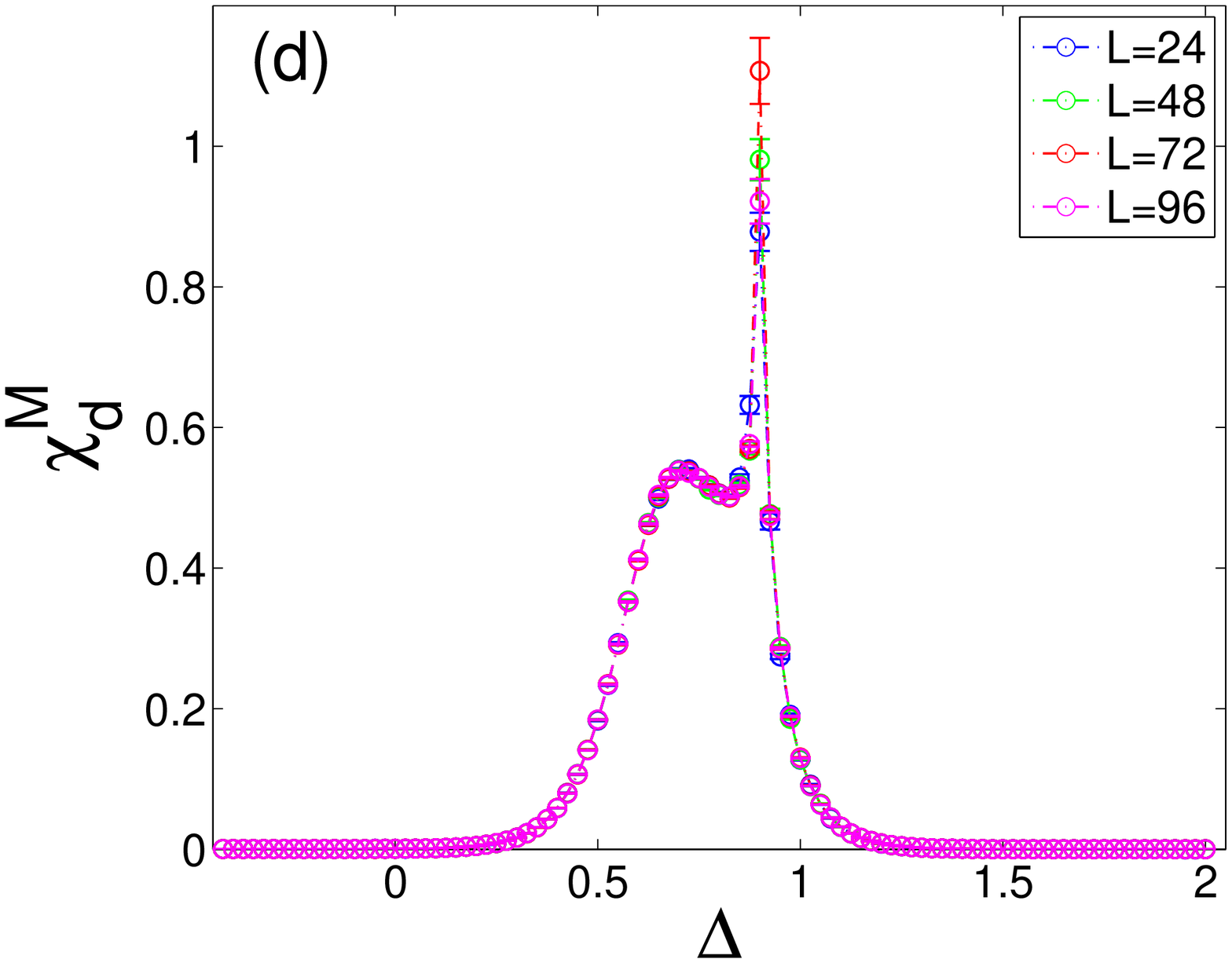}\label{fig:chi-D_J-2_t015}}
\subfigure{\includegraphics[scale=0.25,clip]{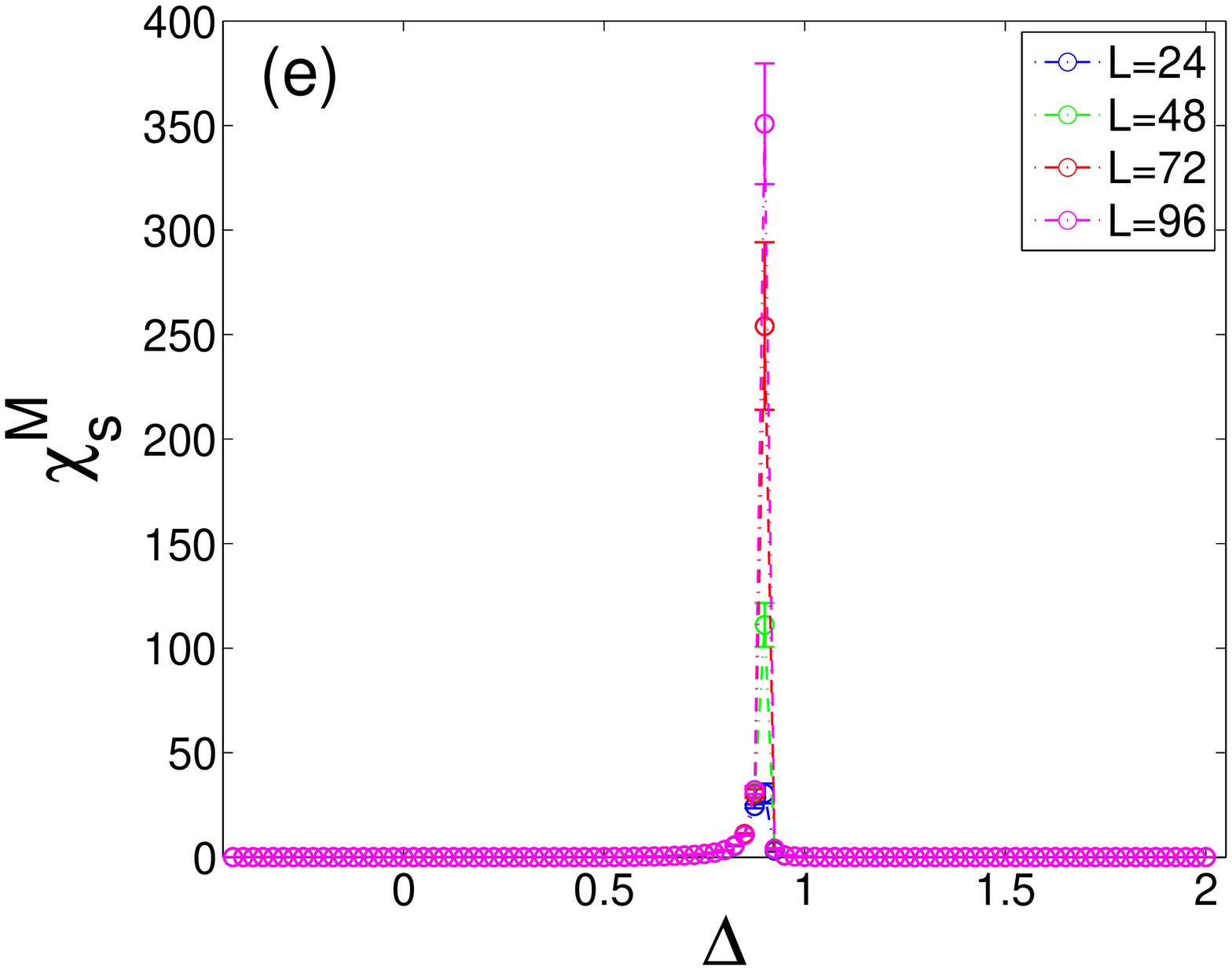}\label{fig:chis-D_J-2_t015}}
\subfigure{\includegraphics[scale=0.25,clip]{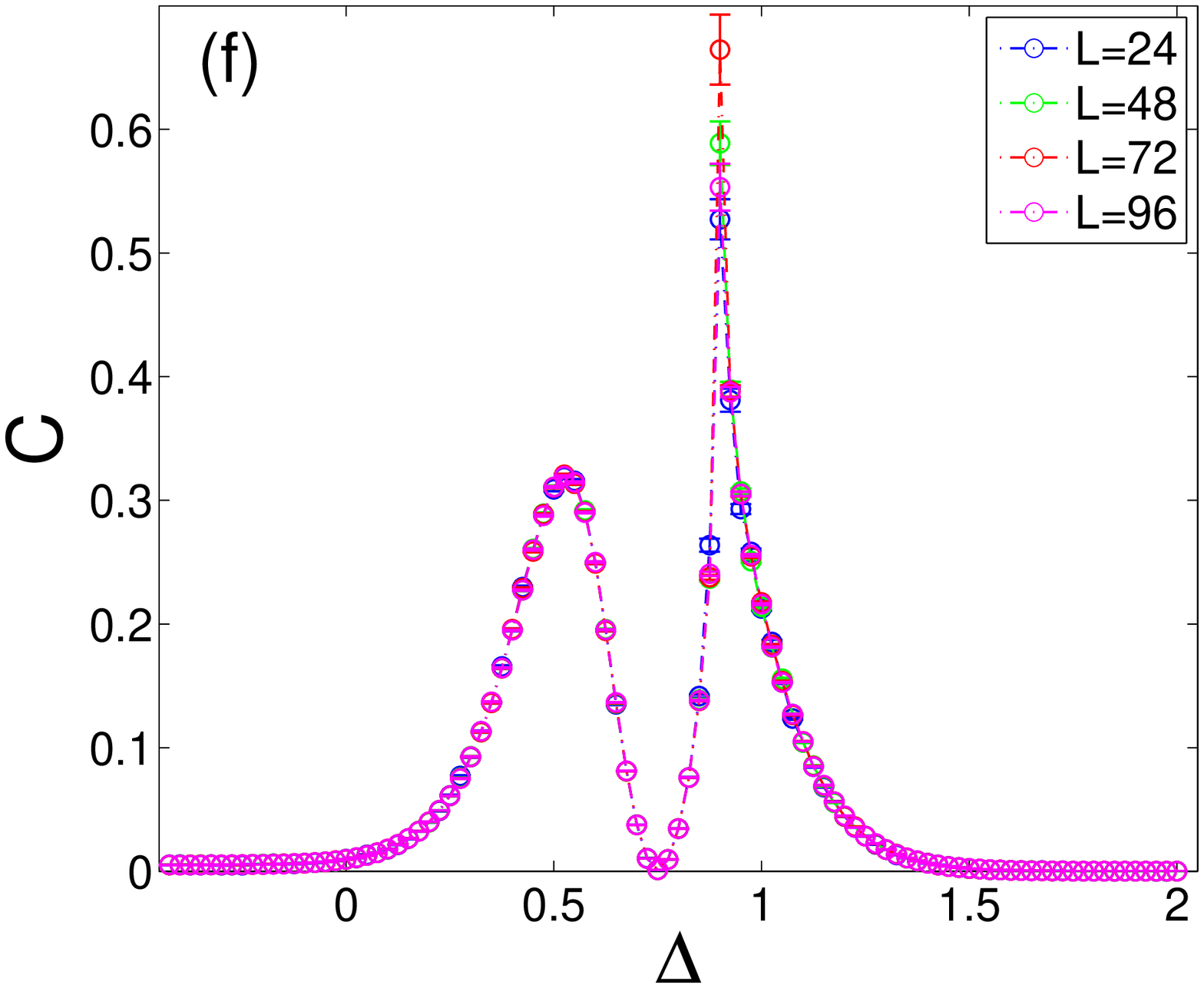}\label{fig:c-D_J-2_t015}}
\caption{$\Delta$-dependence of the quantities around $F_1 \rightarrow FRM$ transition at $t=0.15$.} \label{fig:F1-FRM_t015}
\end{figure}

\begin{figure}[t!]
\centering
\subfigure{\includegraphics[scale=0.25,clip]{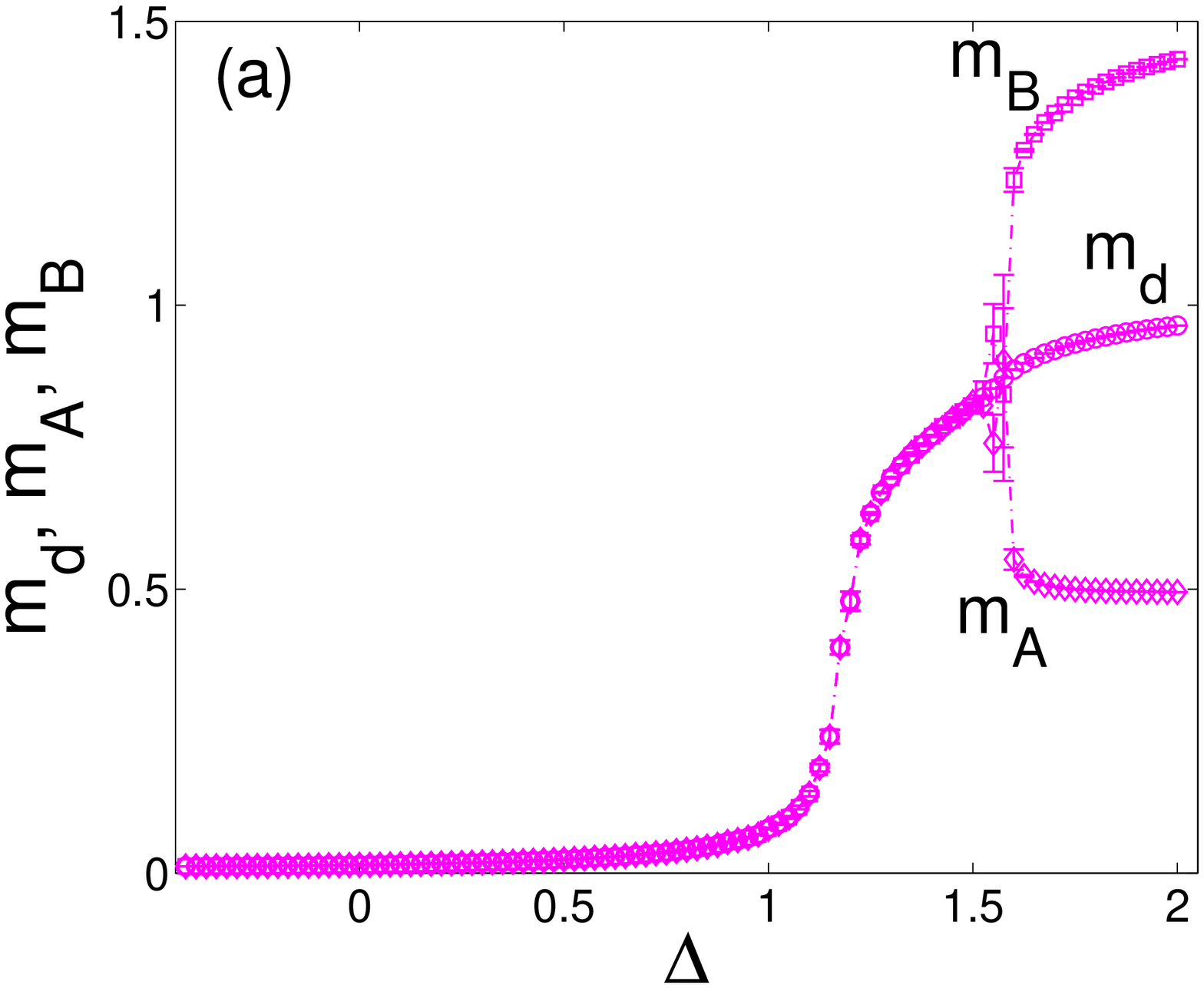}\label{fig:m-D_J-2_t075_PD2}}
\subfigure{\includegraphics[scale=0.25,clip]{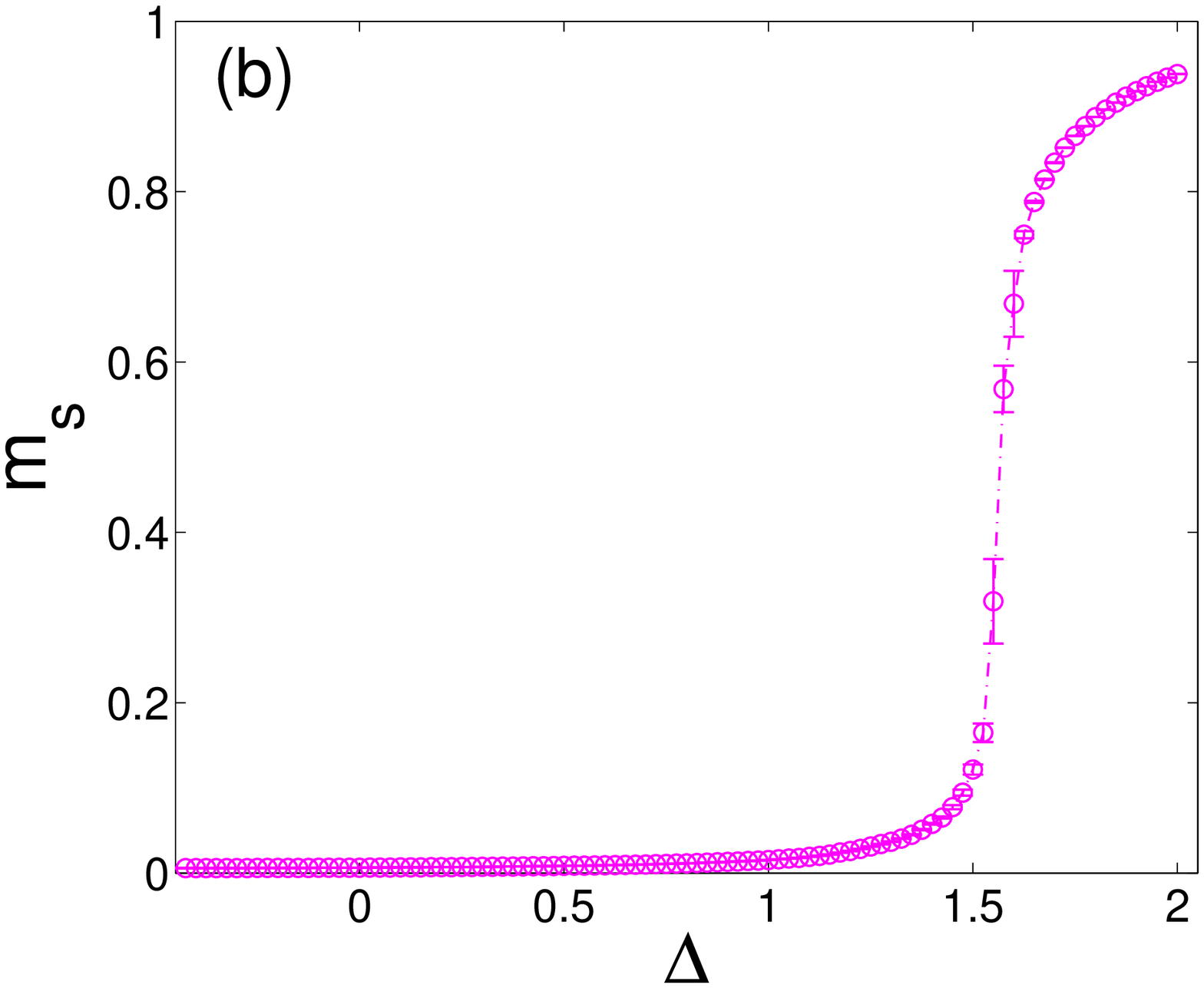}\label{fig:ms-D_J-2_t075_PD2}}
\subfigure{\includegraphics[scale=0.25,clip]{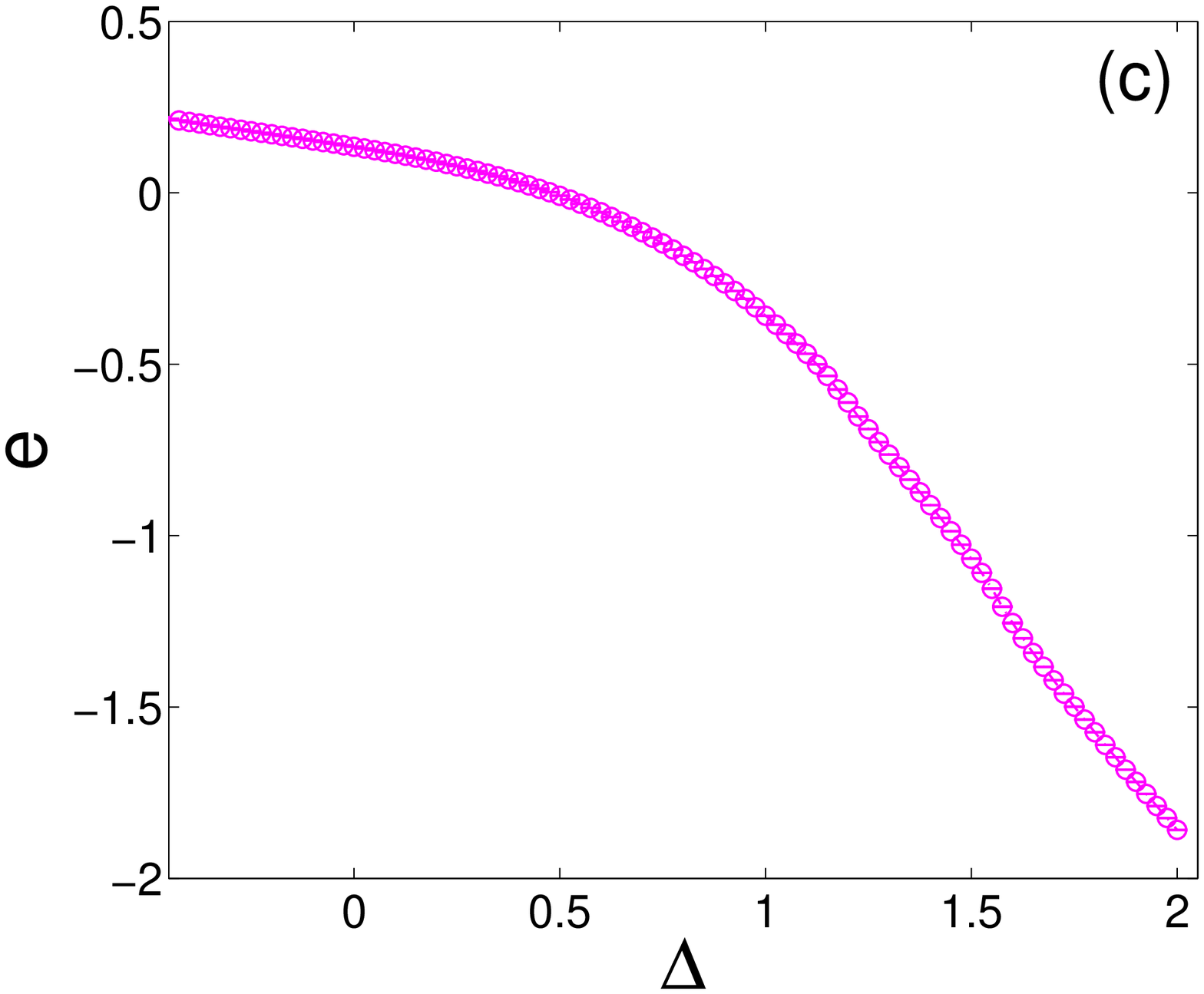}\label{fig:e-D_J-2_t075_PD2}} \\
\subfigure{\includegraphics[scale=0.25,clip]{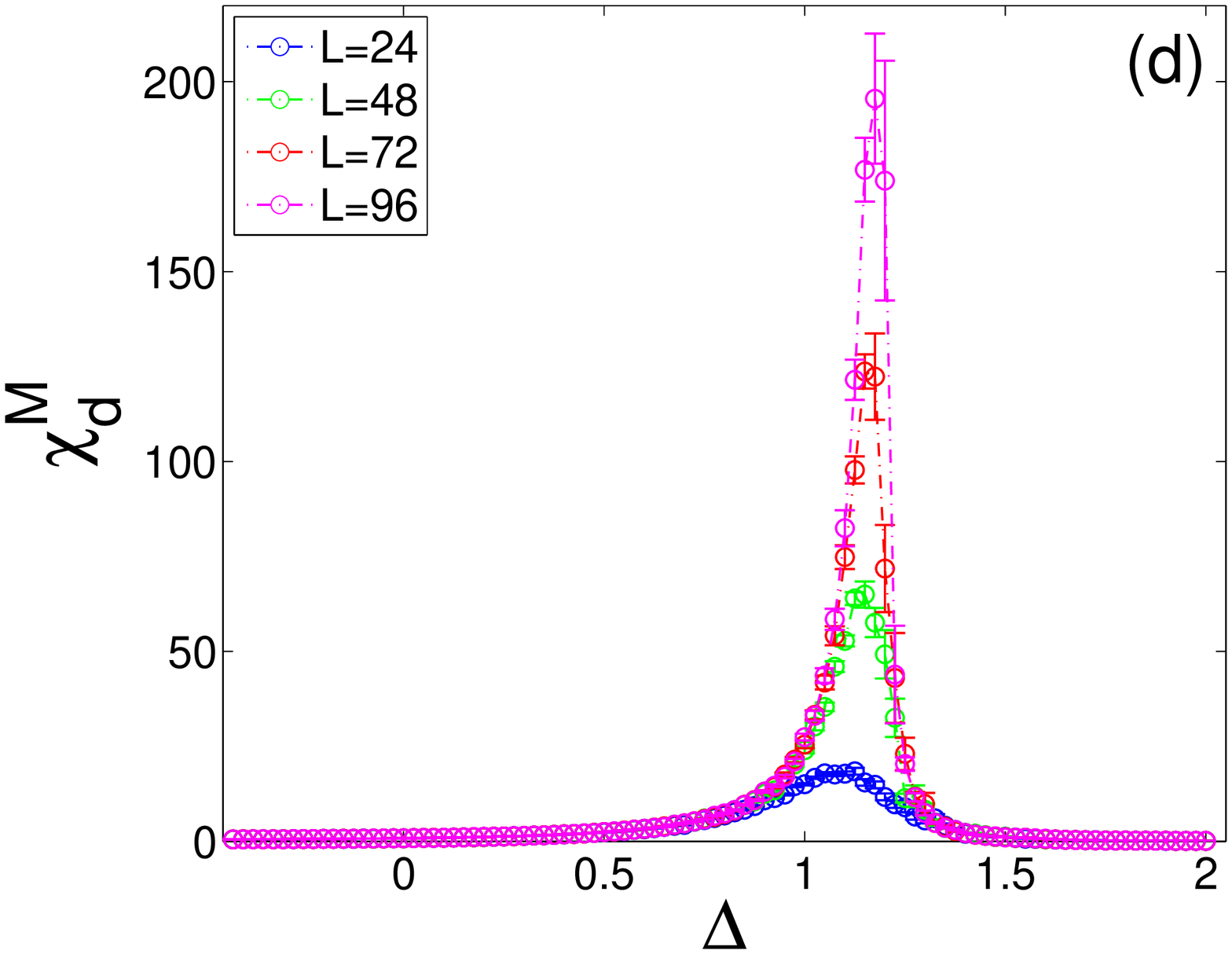}\label{fig:chi-D_J-2_t075_PD2}}
\subfigure{\includegraphics[scale=0.25,clip]{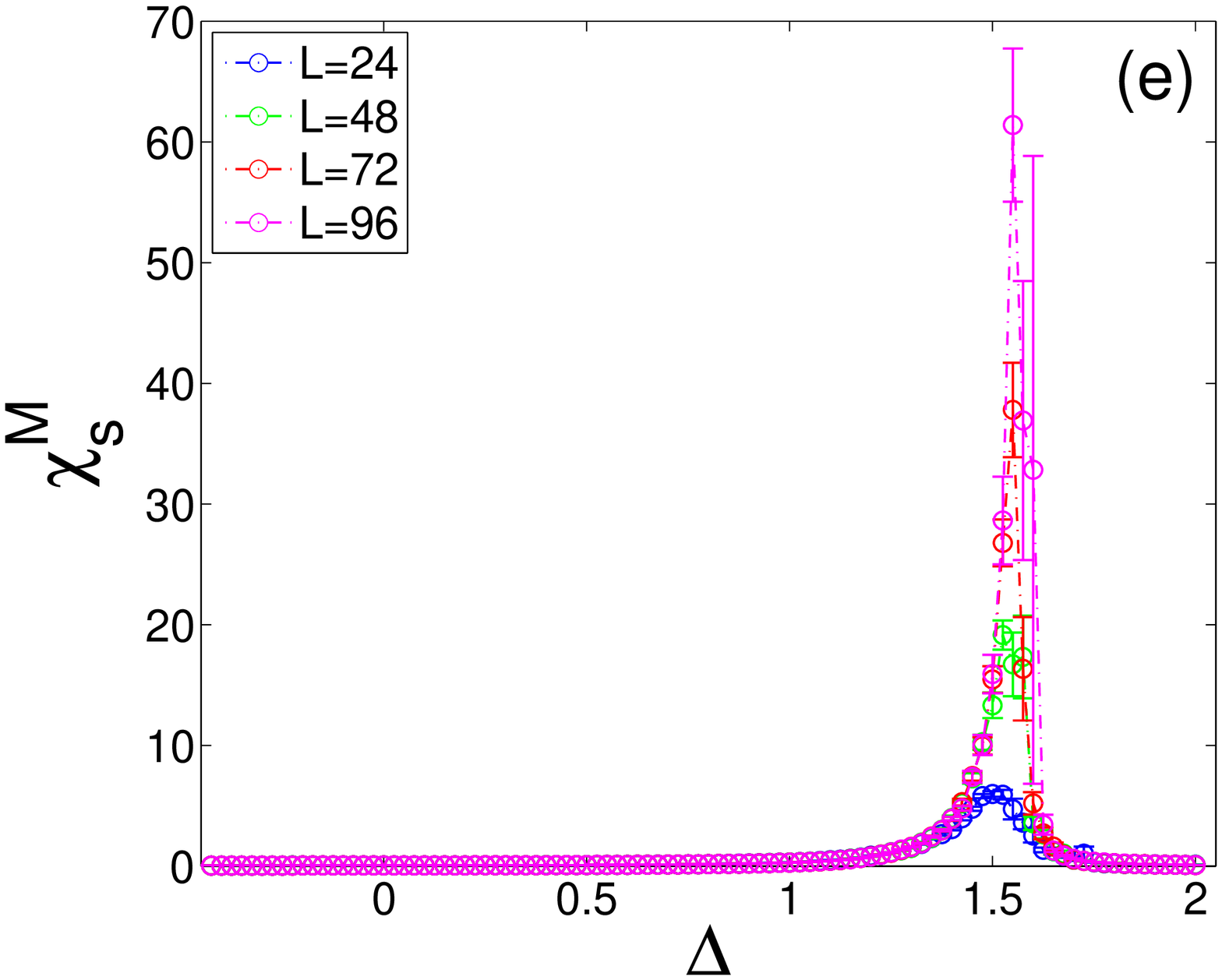}\label{fig:chis-D_J-2_t075_PD2}}
\subfigure{\includegraphics[scale=0.25,clip]{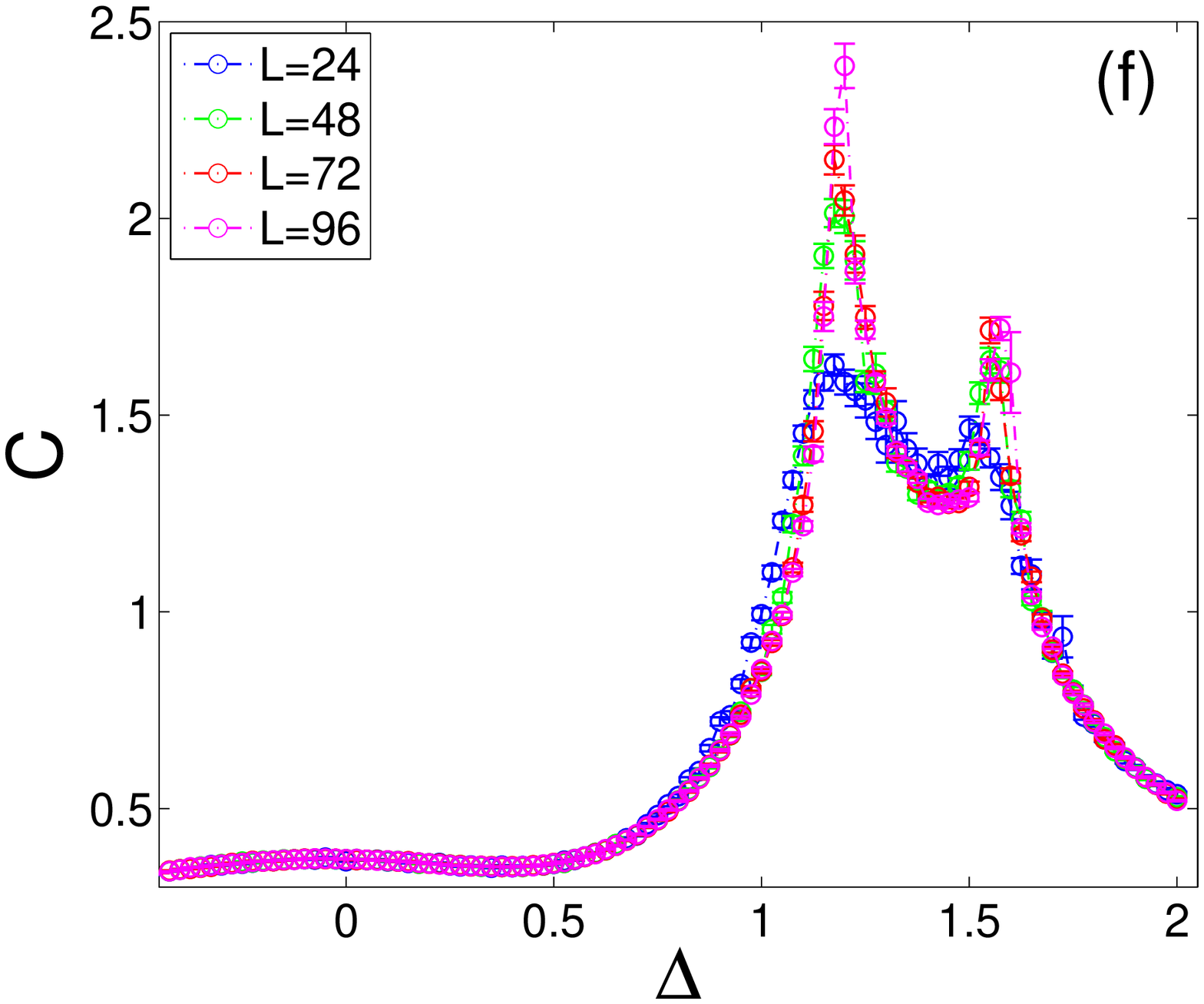}\label{fig:c-D_J-2_t075_PD2}}
\caption{$\Delta$-dependence around $P \rightarrow F_1 \rightarrow FRM$ transition at $t=0.75$.} \label{fig:F1-FRM_t075}
\end{figure}

\begin{figure}[t!]
\centering
\subfigure{\includegraphics[scale=0.25,clip]{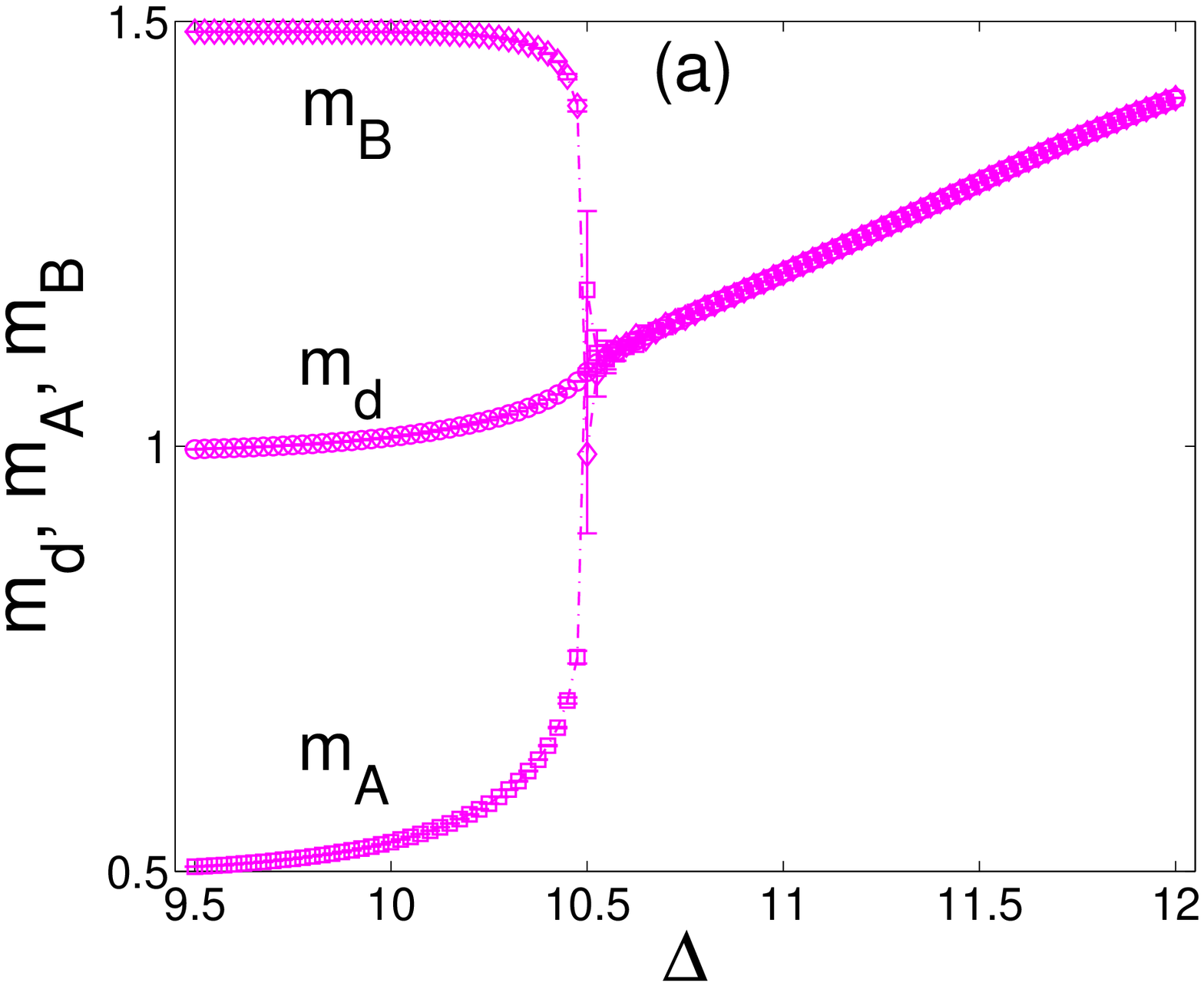}\label{fig:m-D_J-2_t075_D2D3}}
\subfigure{\includegraphics[scale=0.25,clip]{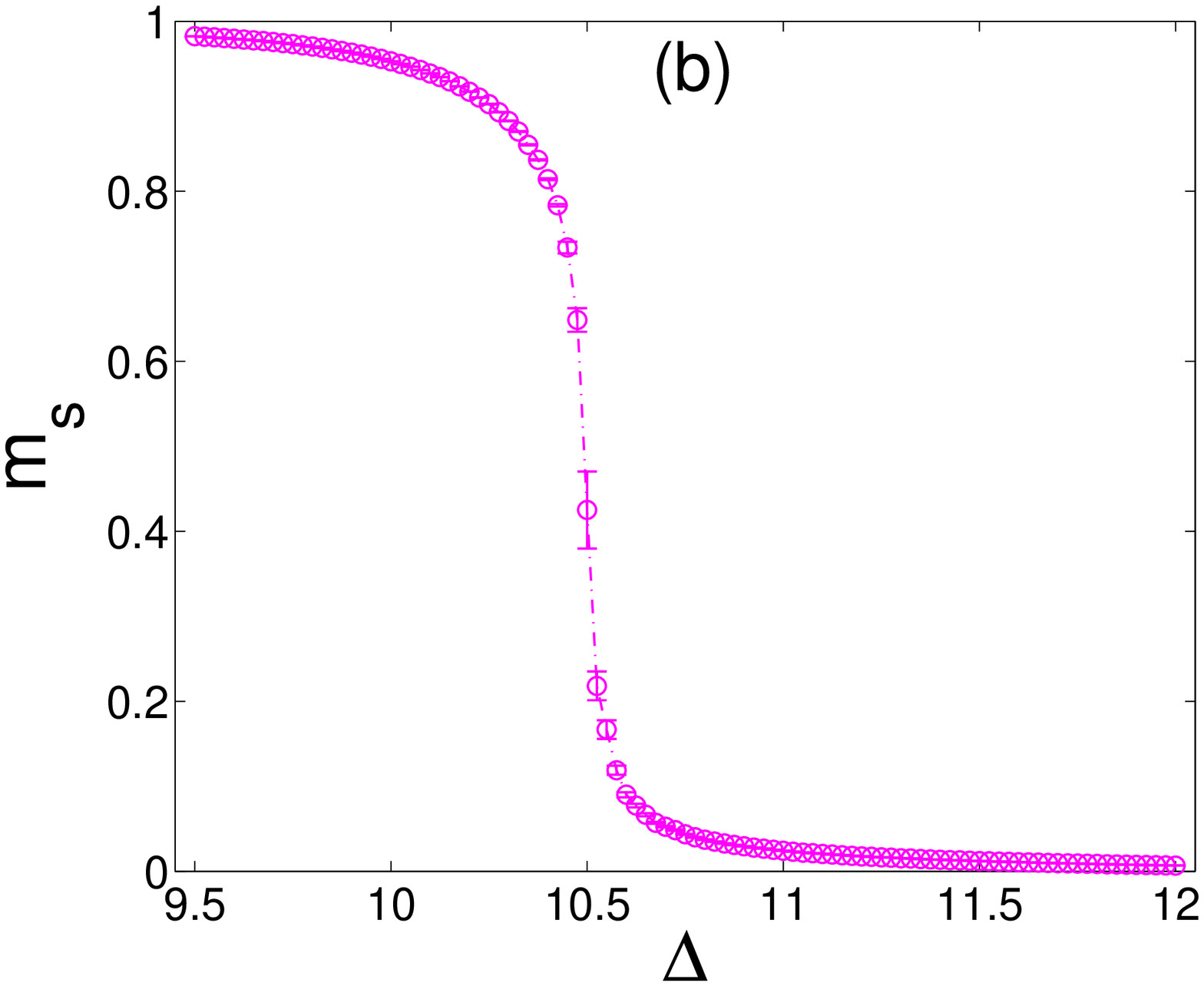}\label{fig:ms-D_J-2_t075_D2D3}}
\subfigure{\includegraphics[scale=0.25,clip]{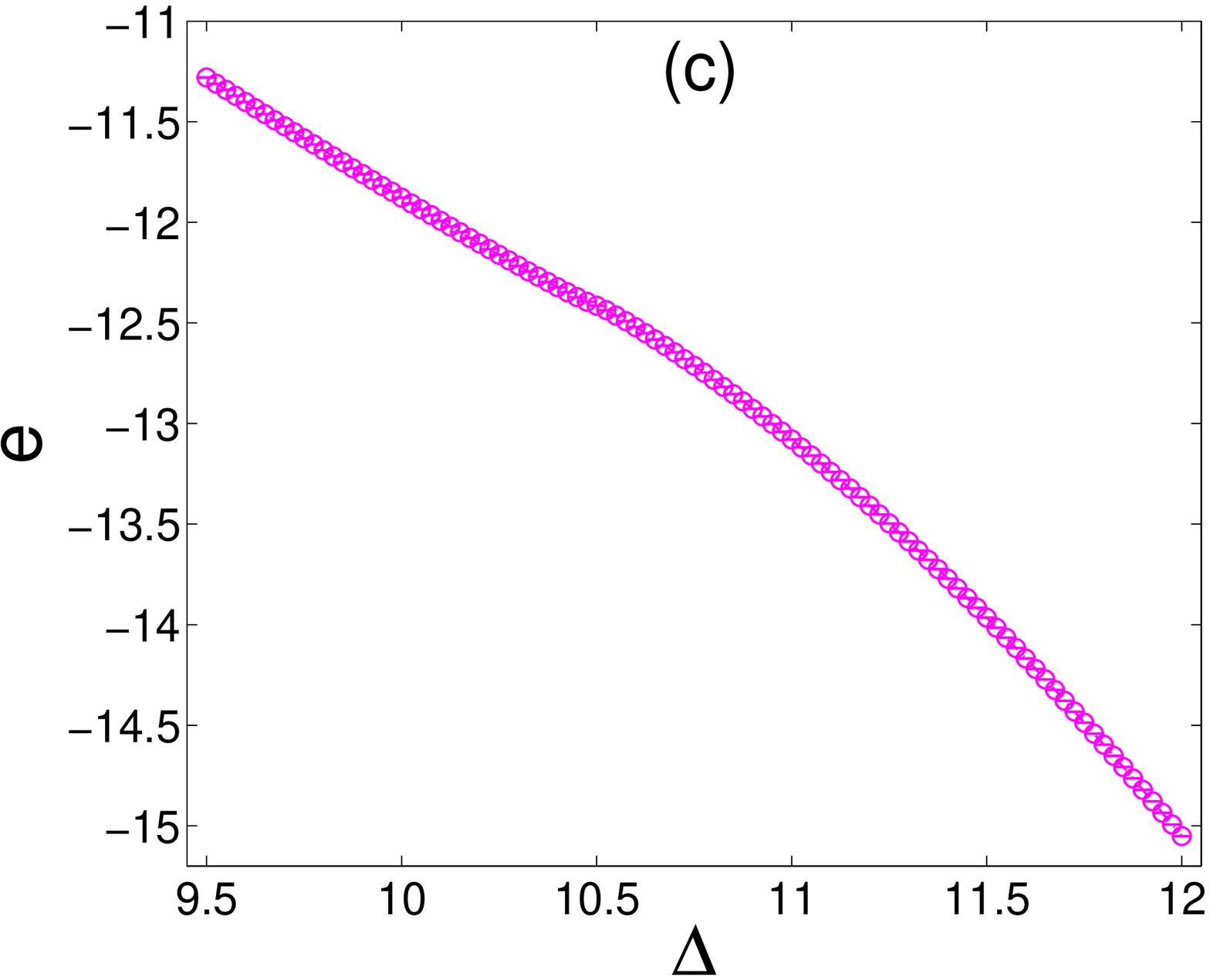}\label{fig:e-D_J-2_t075_D2D3}} \\
\subfigure{\includegraphics[scale=0.25,clip]{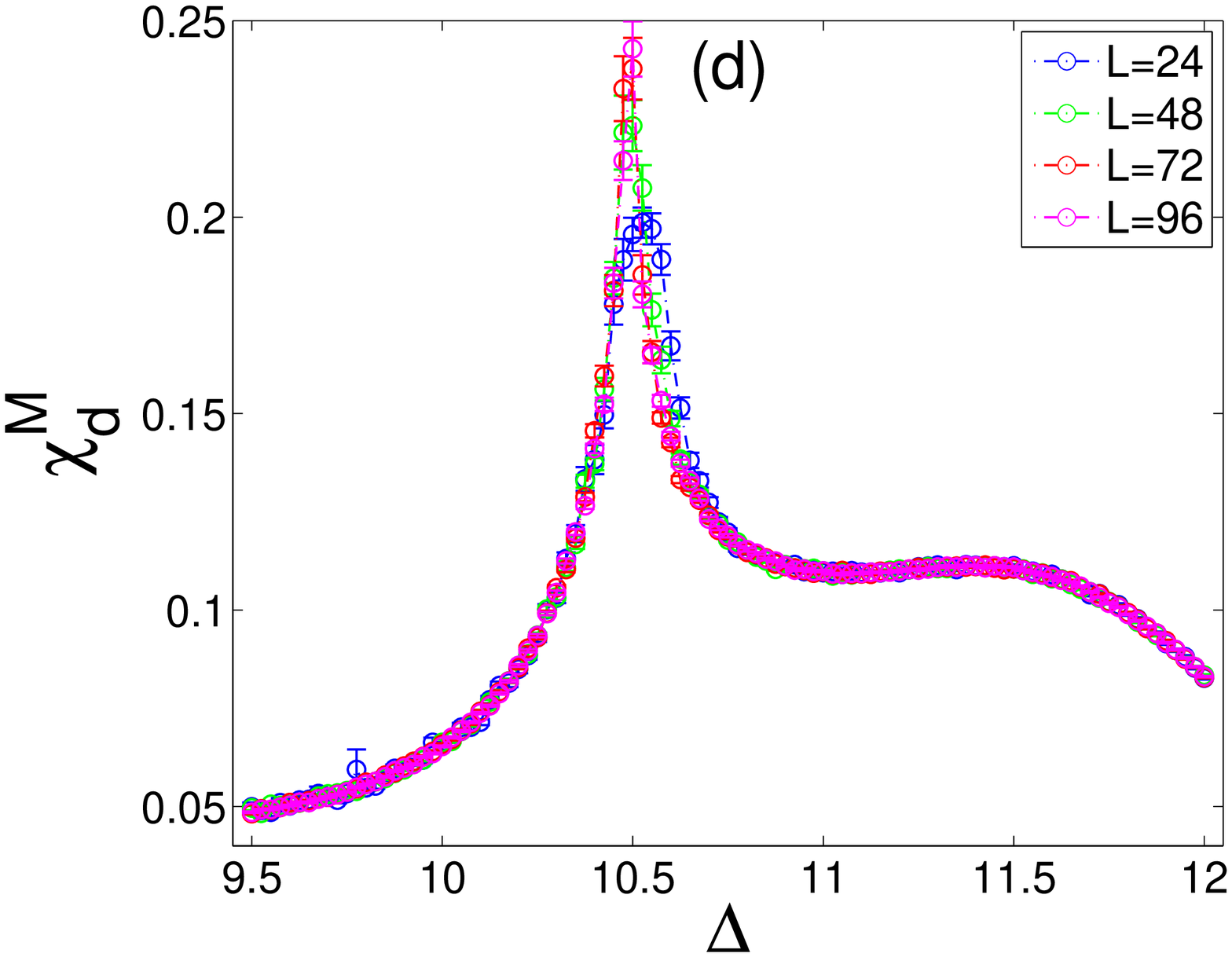}\label{fig:chi-D_J-2_t075_D2D3}}
\subfigure{\includegraphics[scale=0.25,clip]{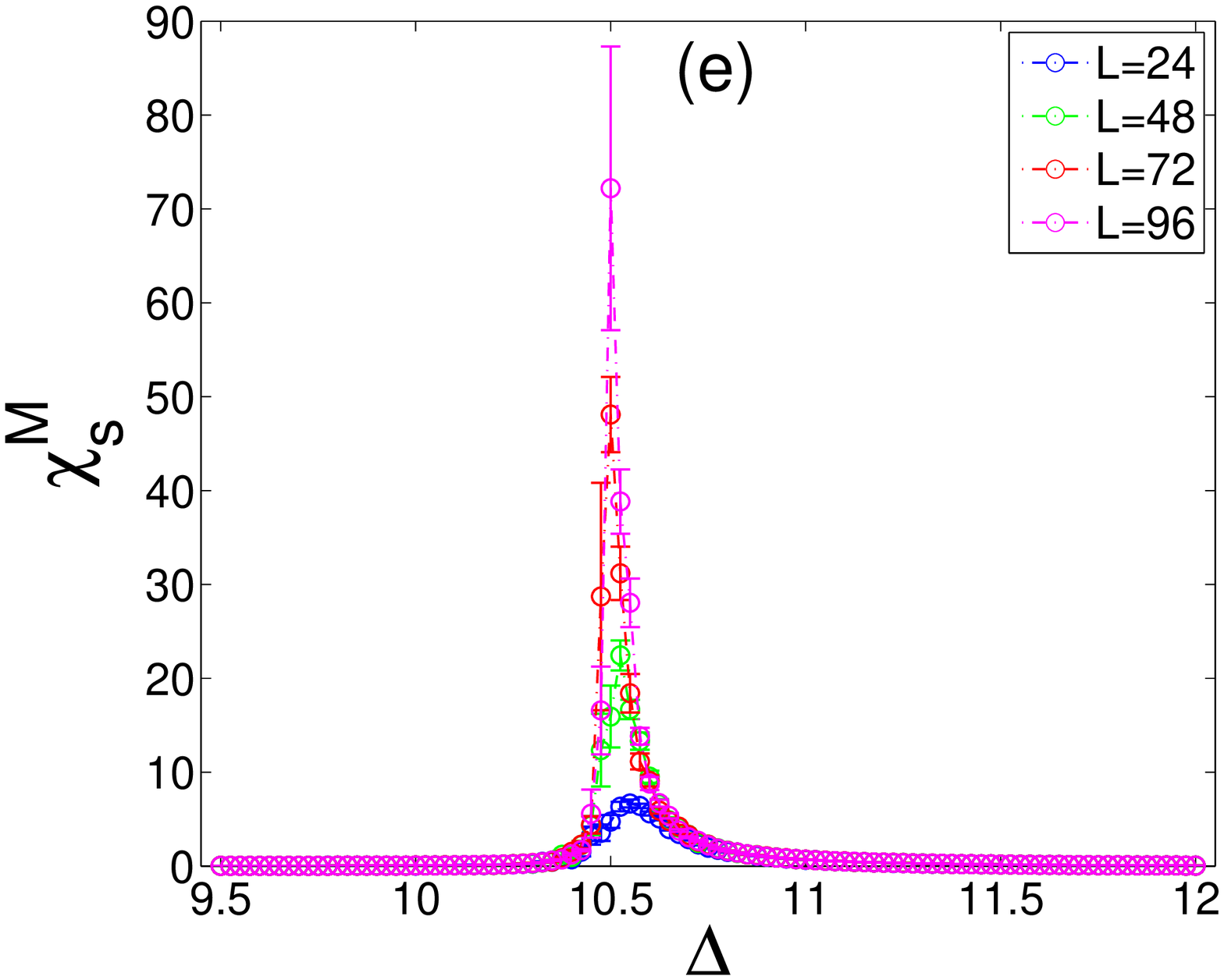}\label{fig:chis-D_J-2_t075_D2D3}}
\subfigure{\includegraphics[scale=0.25,clip]{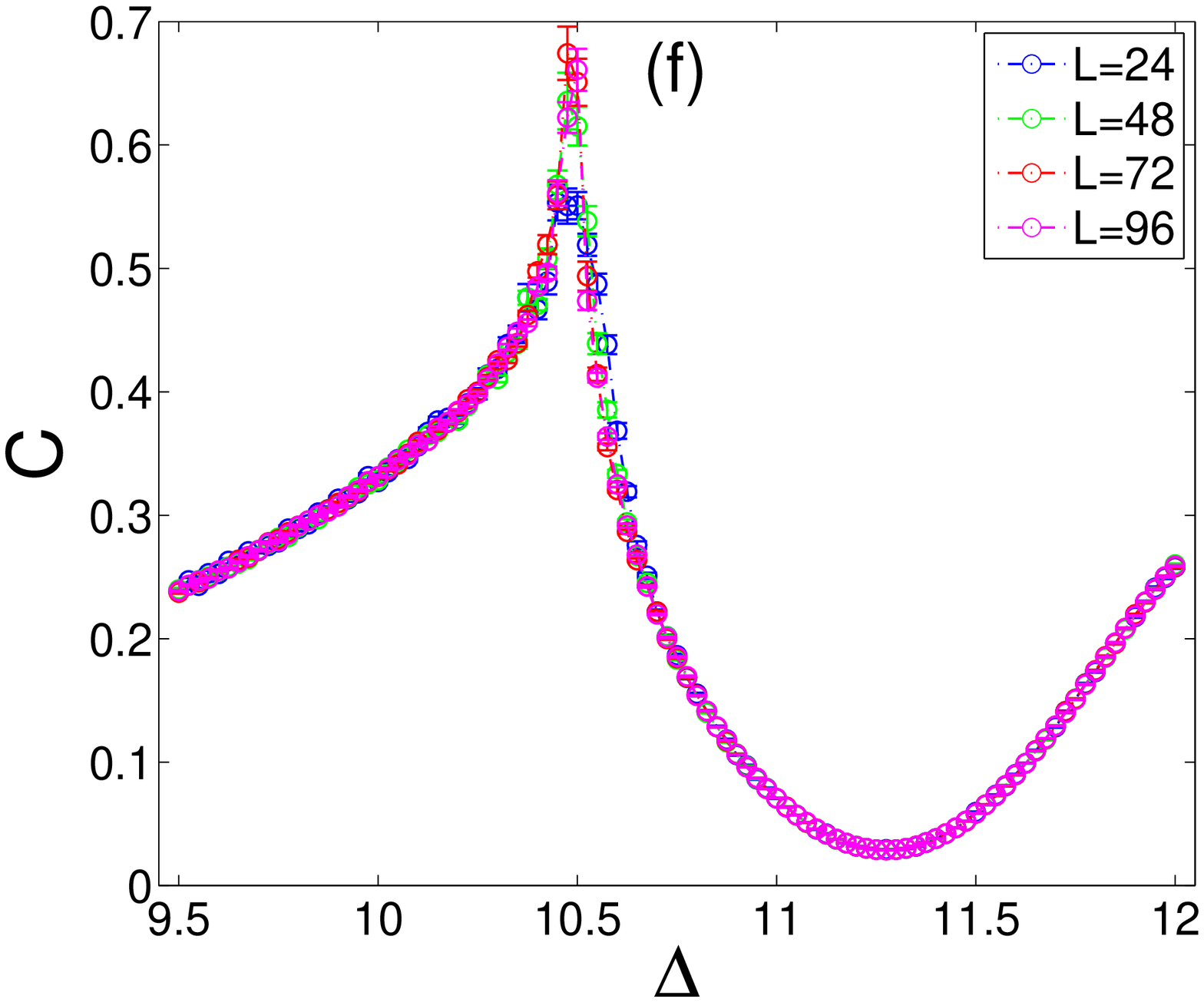}\label{fig:c-D_J-2_t075_D2D3}}
\caption{$\Delta$-dependence around $FRM \rightarrow F_2$ transition at $t=0.75$.} \label{fig:FRM-F2_t075}
\end{figure}

\begin{figure}[]
\centering
\subfigure{\includegraphics[scale=0.25,clip]{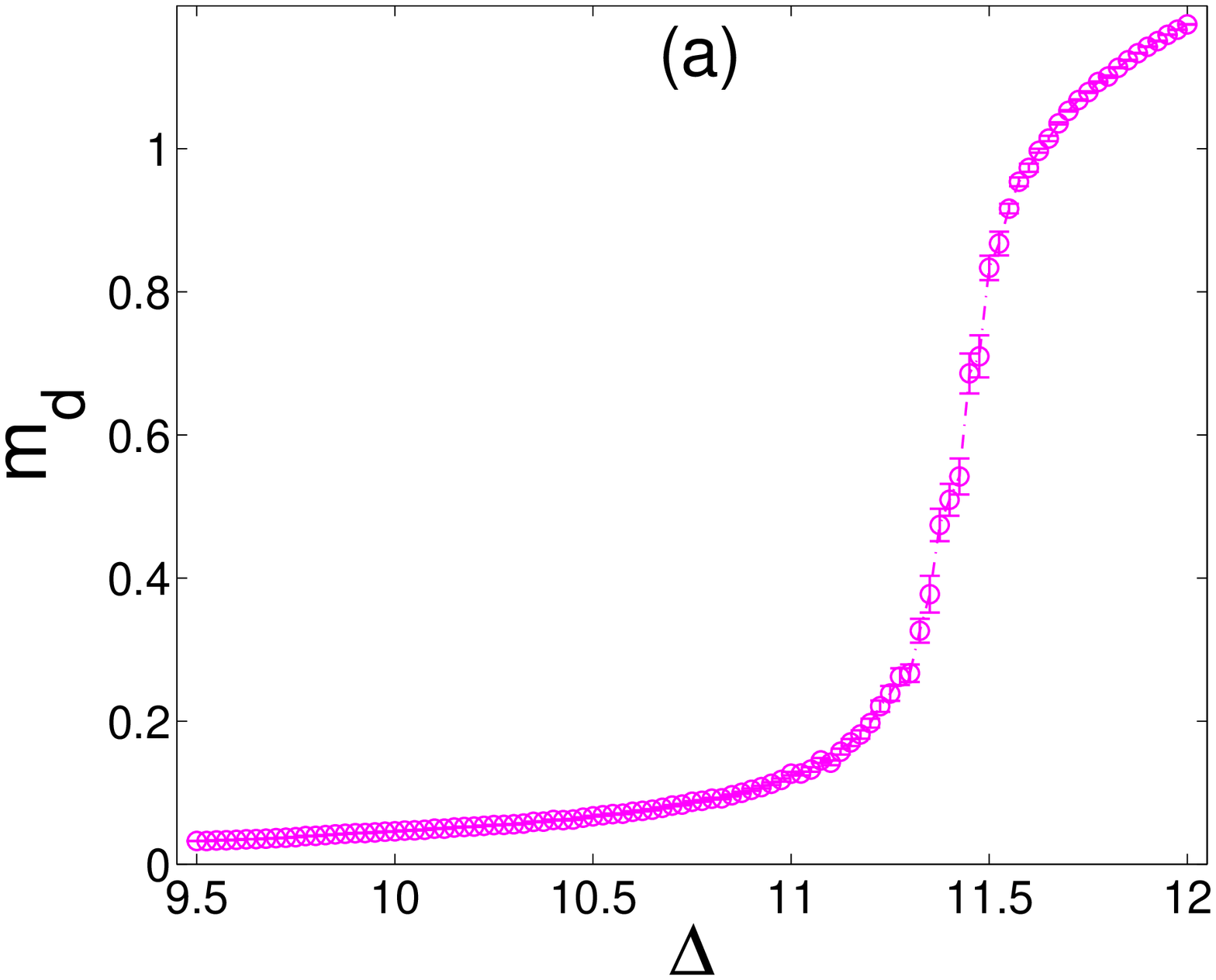}\label{fig:m-D_J-2_t195_QPD3}}
\subfigure{\includegraphics[scale=0.25,clip]{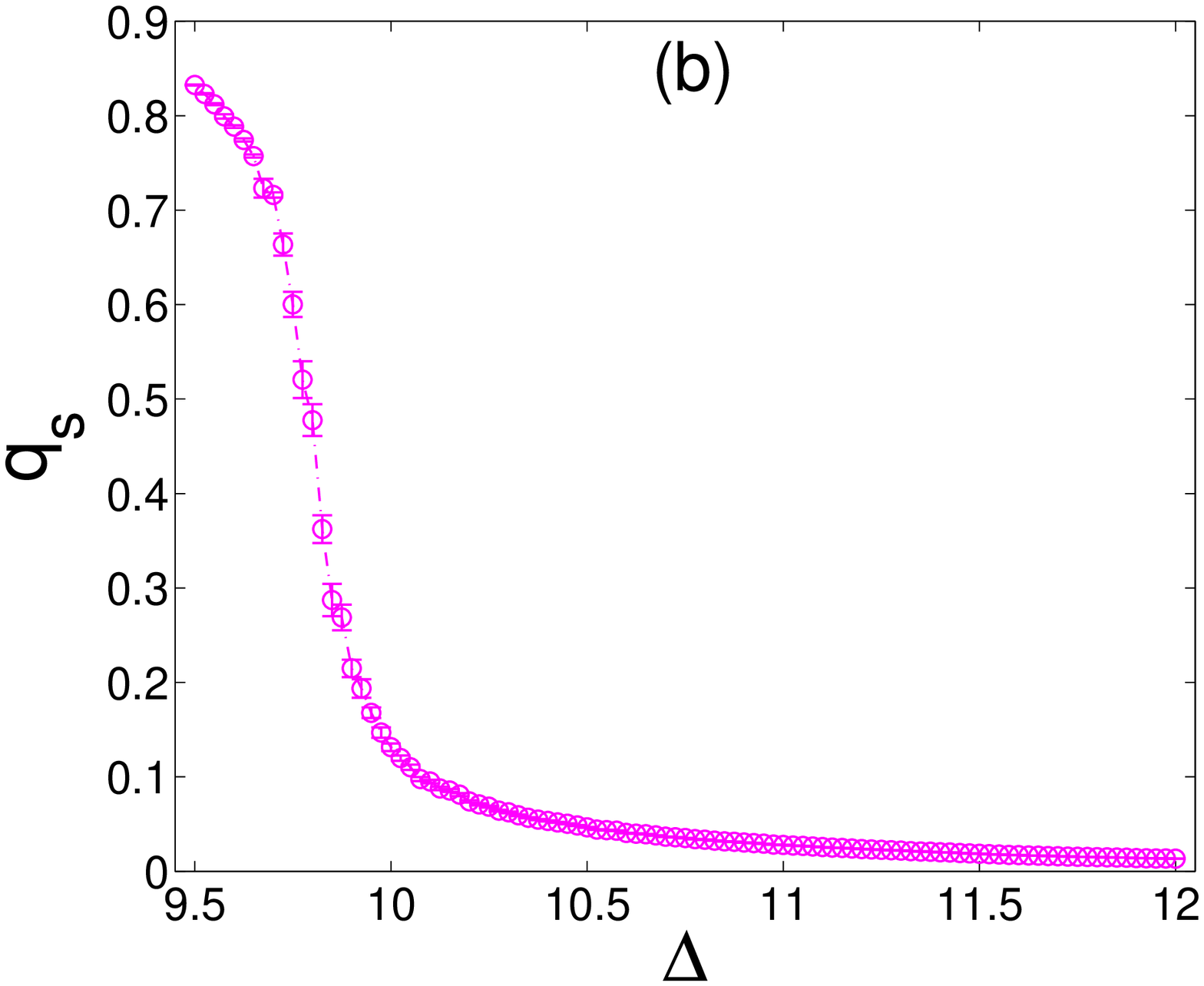}\label{fig:qs-D_J-2_t195_QPD3}}
\subfigure{\includegraphics[scale=0.25,clip]{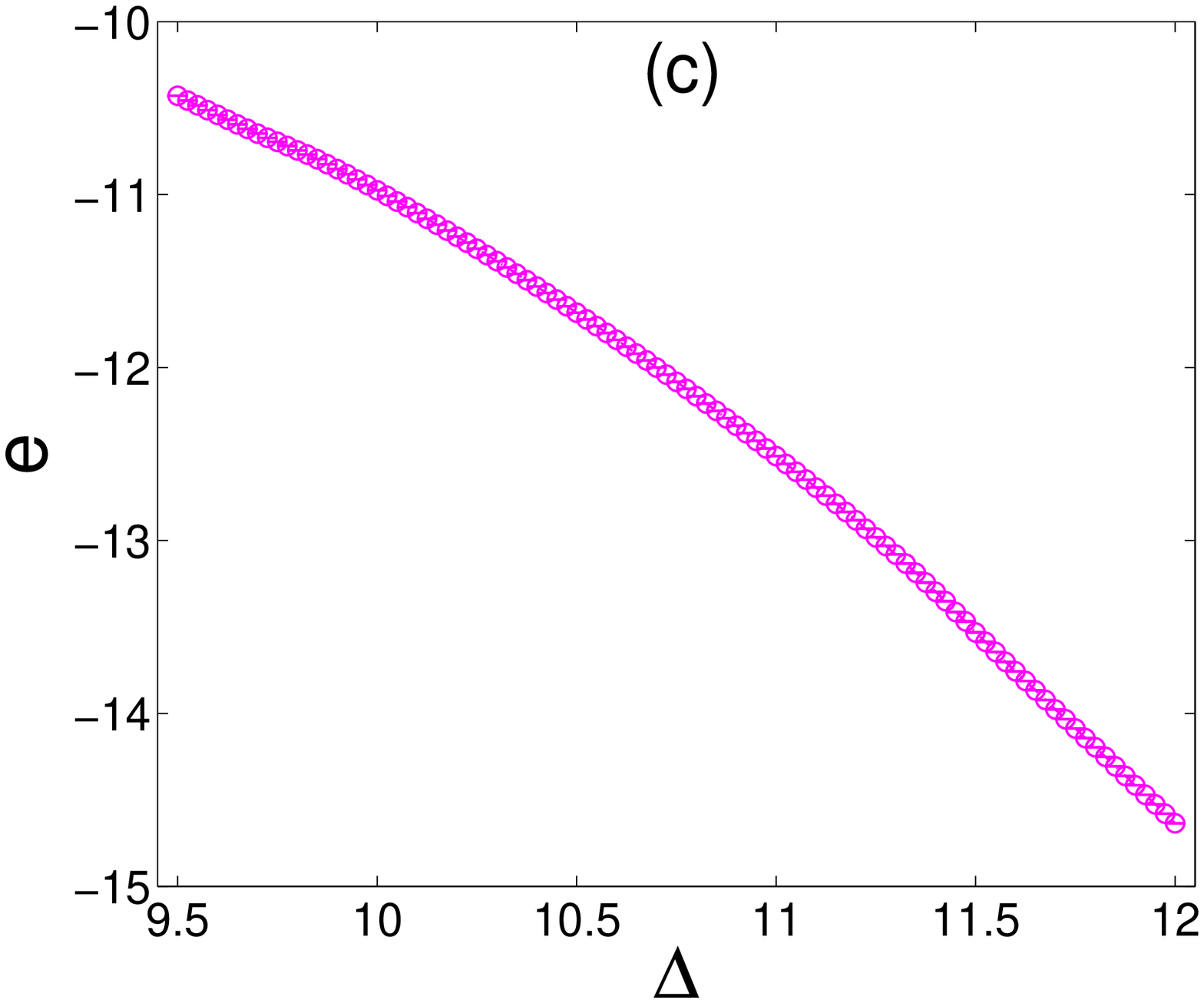}\label{fig:e-D_J-2_t195_QPD3}} \\
\subfigure{\includegraphics[scale=0.25,clip]{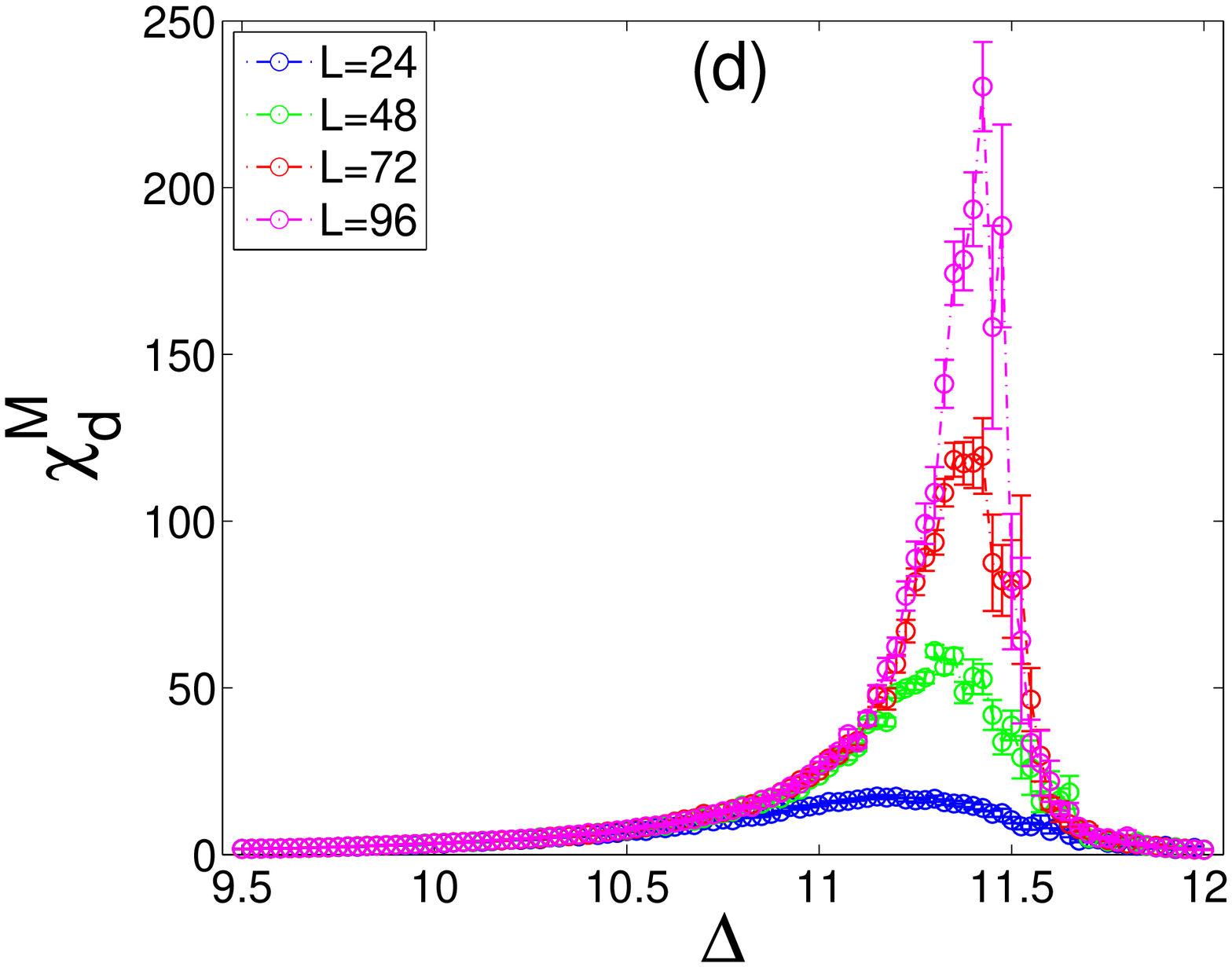}\label{fig:chi-D_J-2_t195_QPD3}}
\subfigure{\includegraphics[scale=0.25,clip]{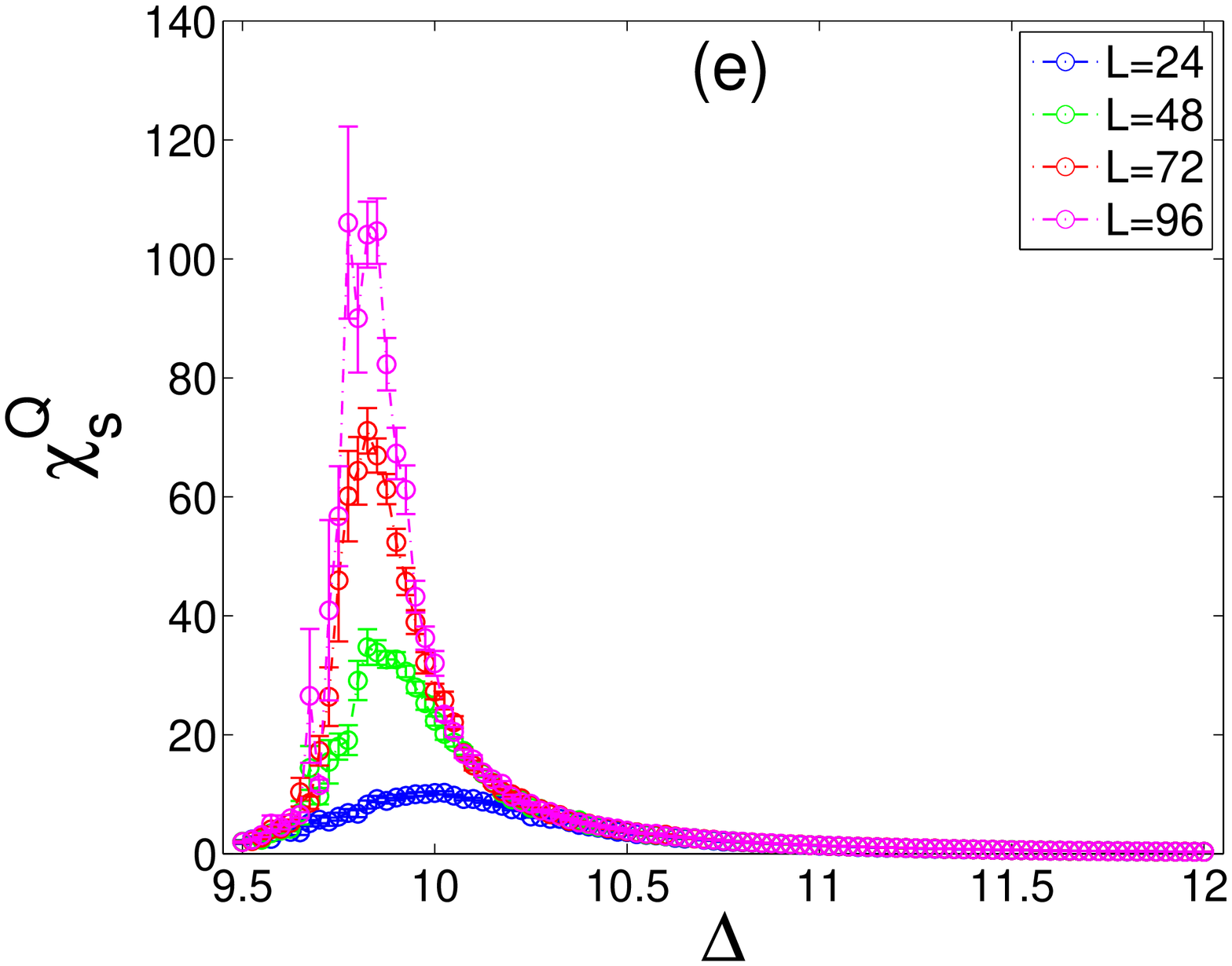}\label{fig:chiqs-D_J-2_t195_QPD3}}
\subfigure{\includegraphics[scale=0.25,clip]{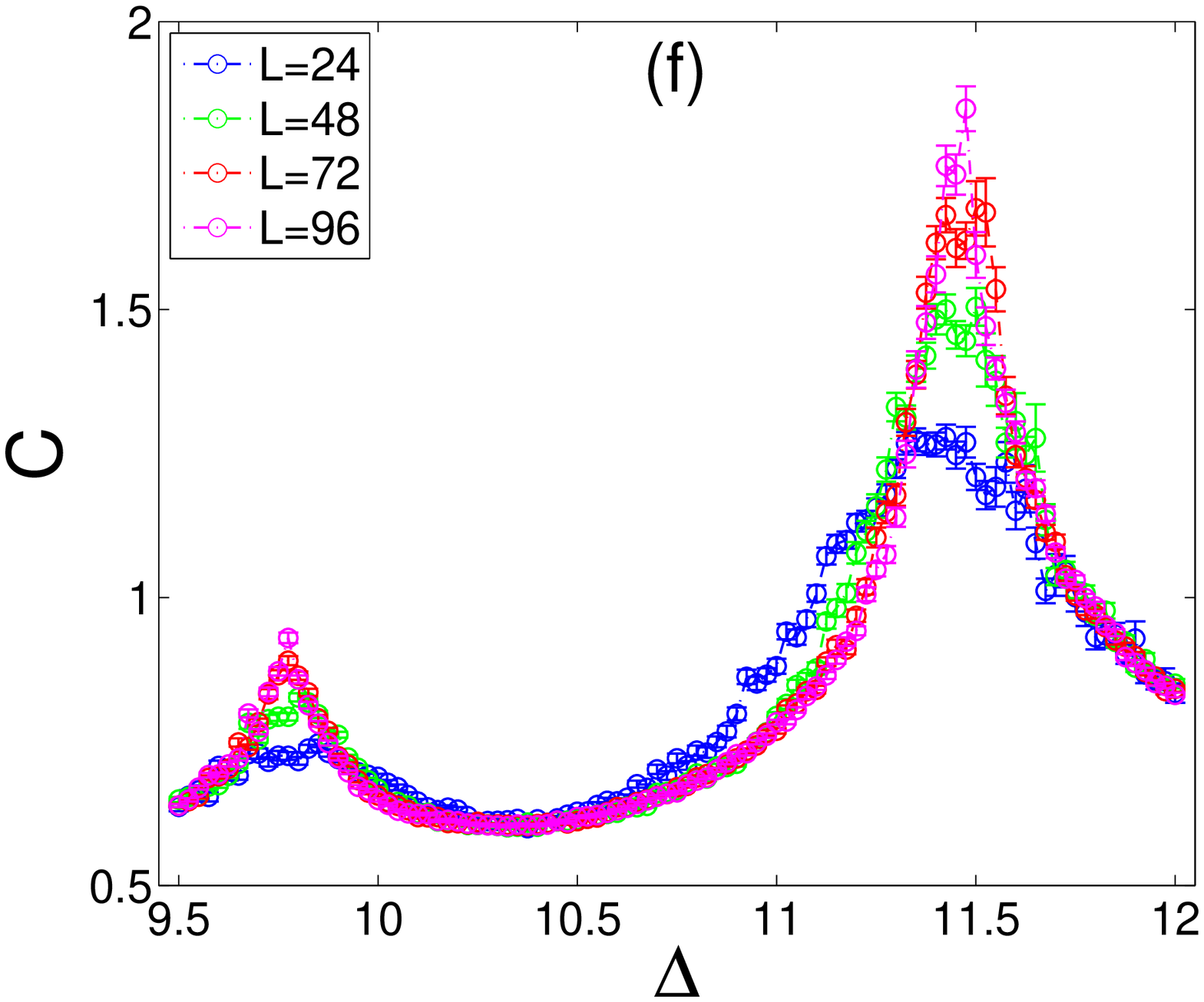}\label{fig:c-D_J-2_t195_QPD3}}
\caption{$\Delta$-dependence of the quantities around $FRQ \rightarrow P \rightarrow F_2$ transition at $t=1.95$.} \label{fig:FRQ-P-F2_t195}
\end{figure}

\begin{figure}[t!]
\centering
\subfigure{\includegraphics[scale=0.28,clip]{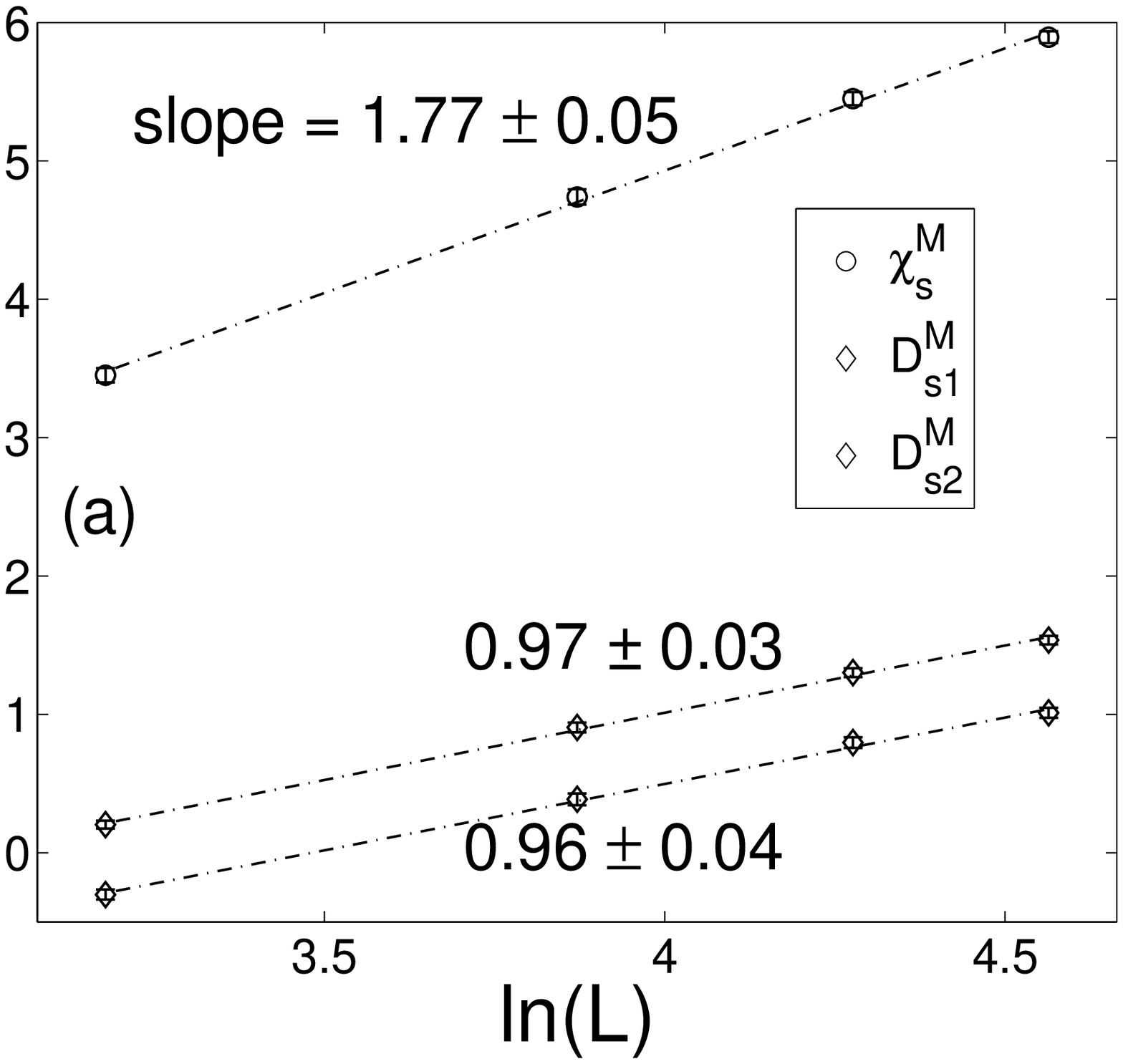}\label{fss_T0_15_D1-D2}}
\subfigure{\includegraphics[scale=0.28,clip]{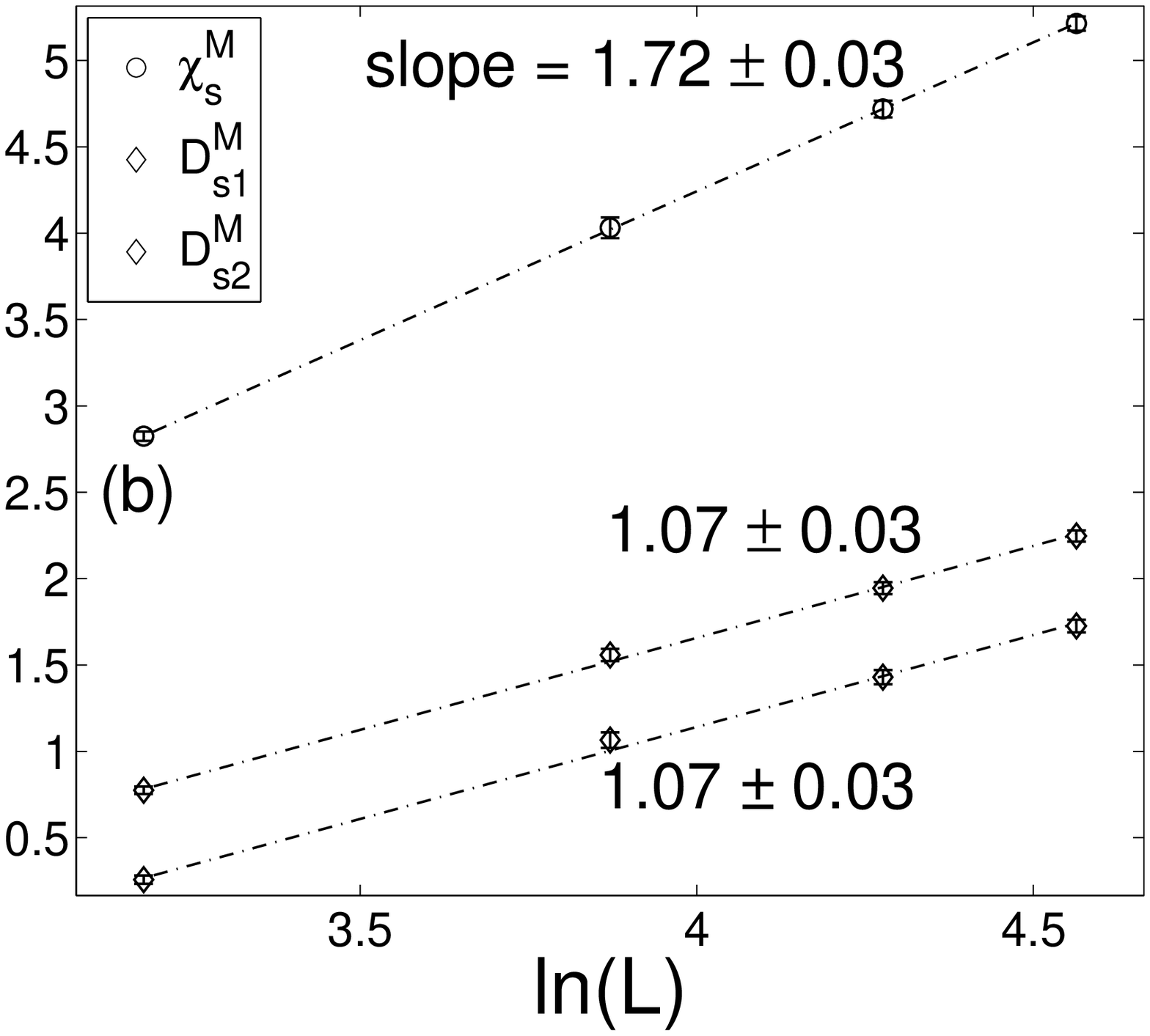}\label{fss_T0_3_D1-D2}}
\subfigure{\includegraphics[scale=0.28,clip]{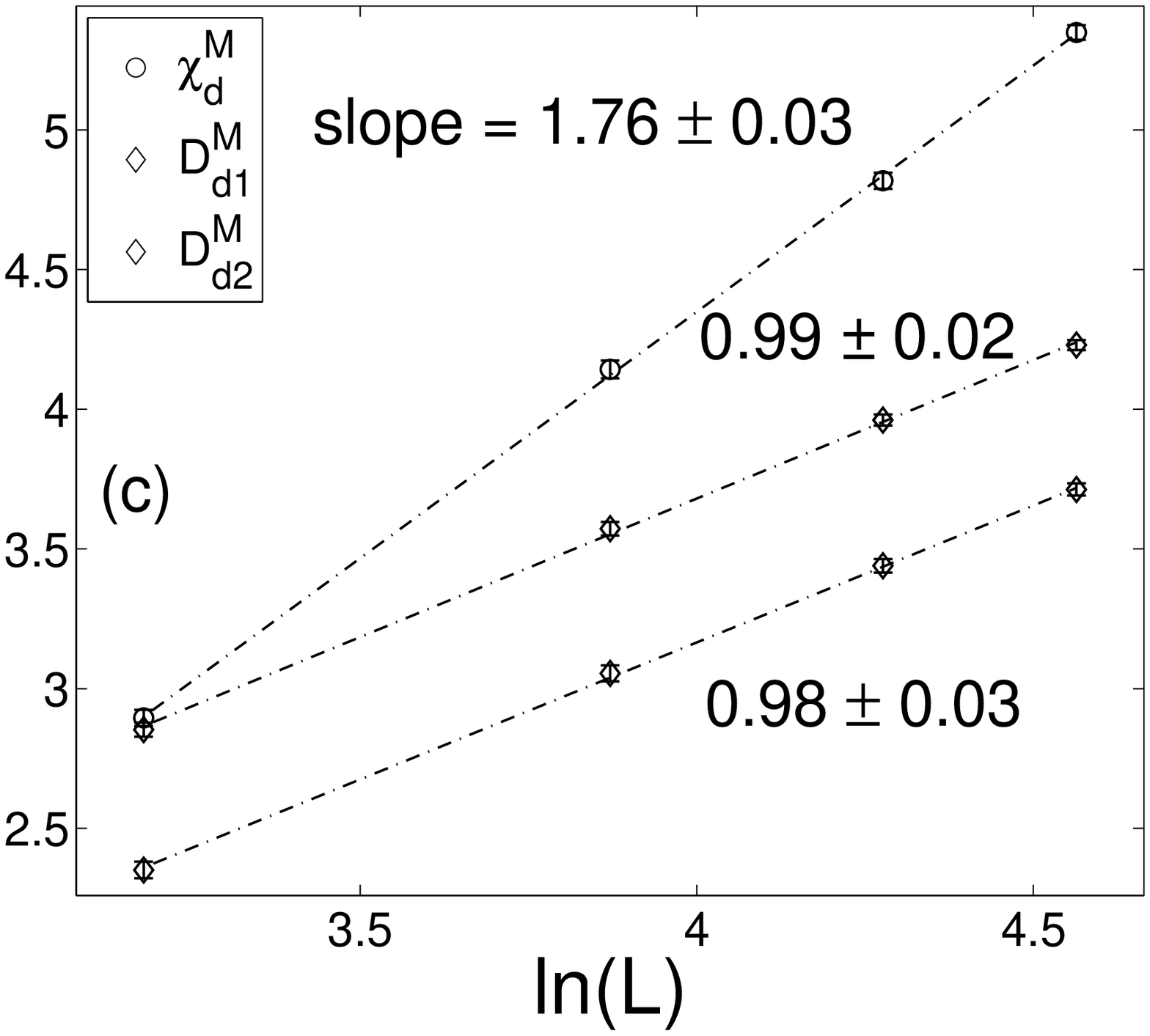}\label{fss_D5_Q-D2_ma}} \\
\subfigure{\includegraphics[scale=0.28,clip]{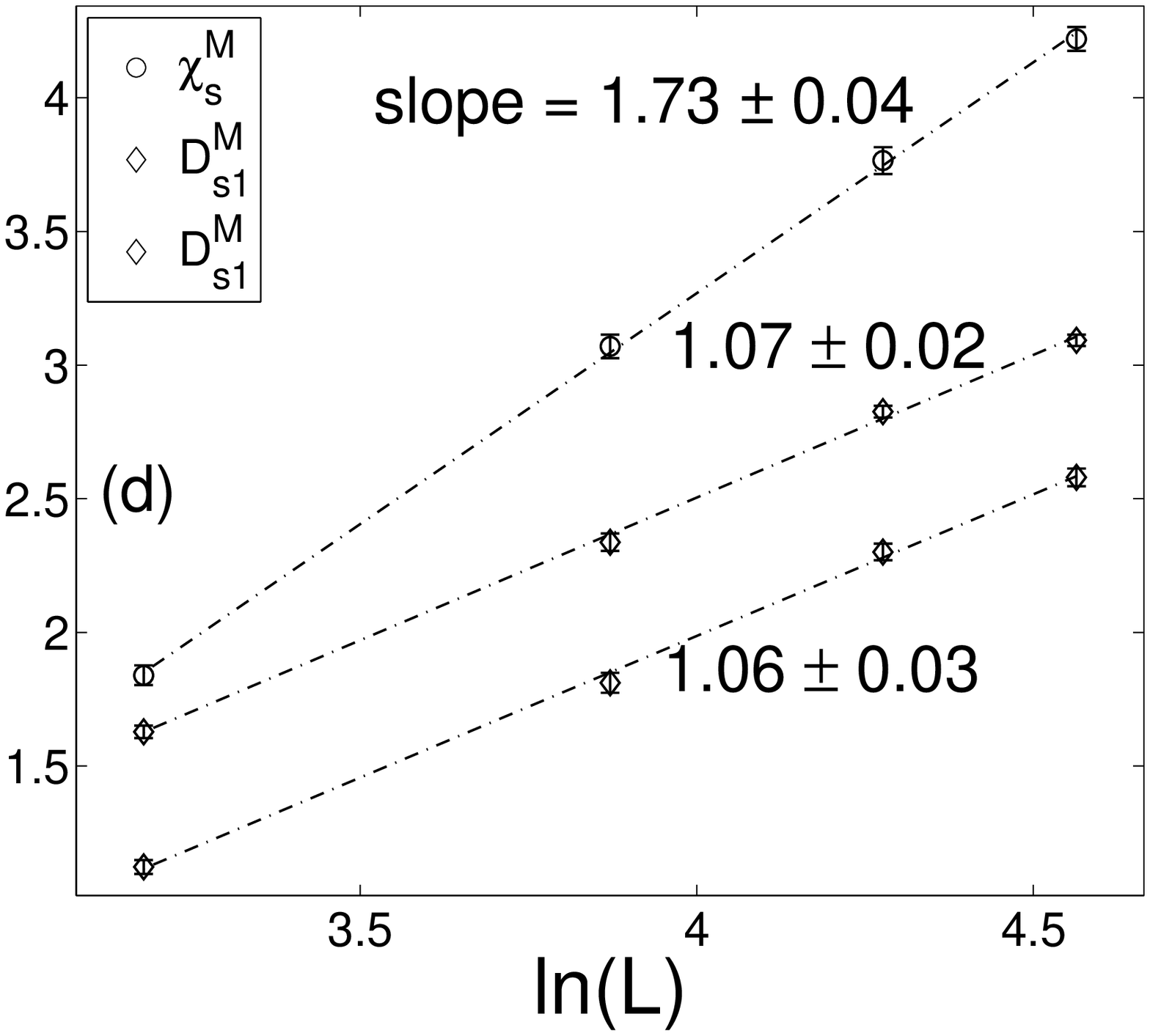}\label{fss_T0_75_D2-D3}}
\subfigure{\includegraphics[scale=0.28,clip]{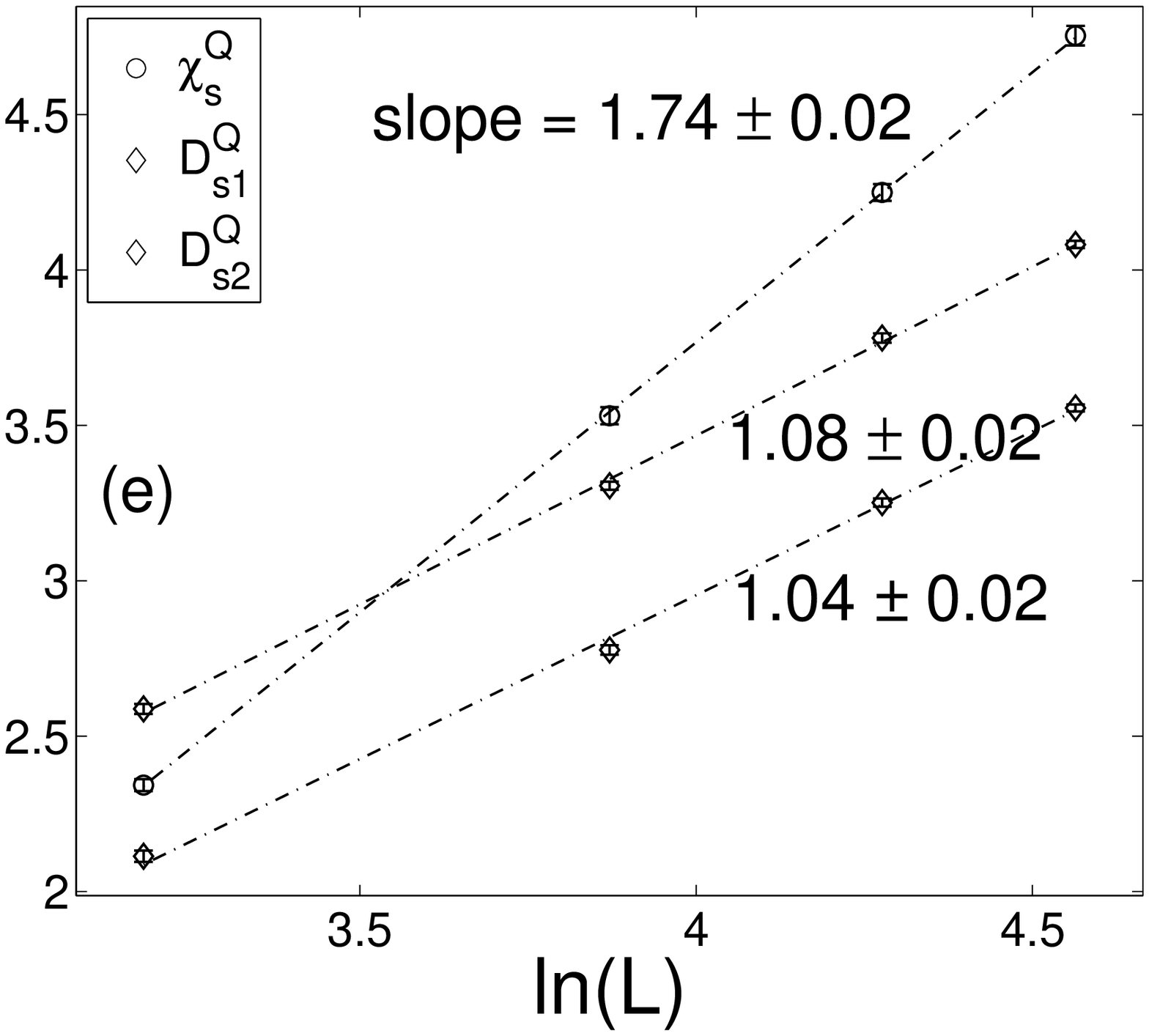}\label{fss_T1_95_P-Q}}
\subfigure{\includegraphics[scale=0.28,clip]{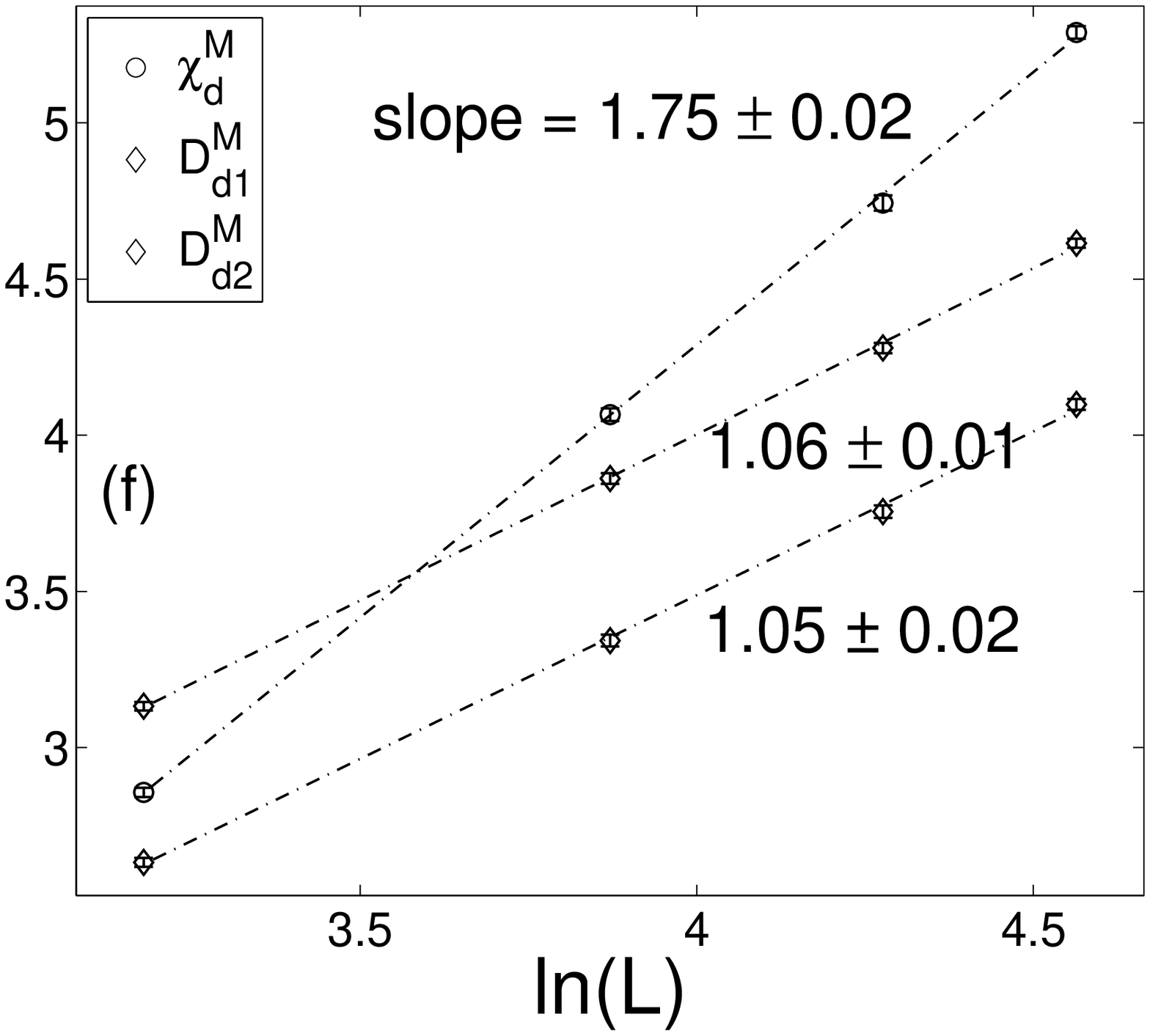}\label{fss_T1_95_P-D3}}
\caption{Critical exponents' ratios $\gamma/\nu$ and $1/\nu$ from the scaling relations (\ref{scalchi}-\ref{scalD2}) for the phase transitions: (a) $F_1 \rightarrow F_2$ at $t=0.15$, (b) $F_1 \rightarrow F_2$ at $t=0.3$, (c) $FRQ \rightarrow FRM$ at $\Delta=5$, (d) $FRM \rightarrow F_2$ at $t=0.75$, (e) $FRQ \rightarrow P$ at $t=1.95$, (f) $P \rightarrow F_2$ at $t=1.95$.}\label{fig:fss_uni}
\end{figure}

\begin{figure}[t!]
\centering
\subfigure{\includegraphics[scale=0.28,clip]{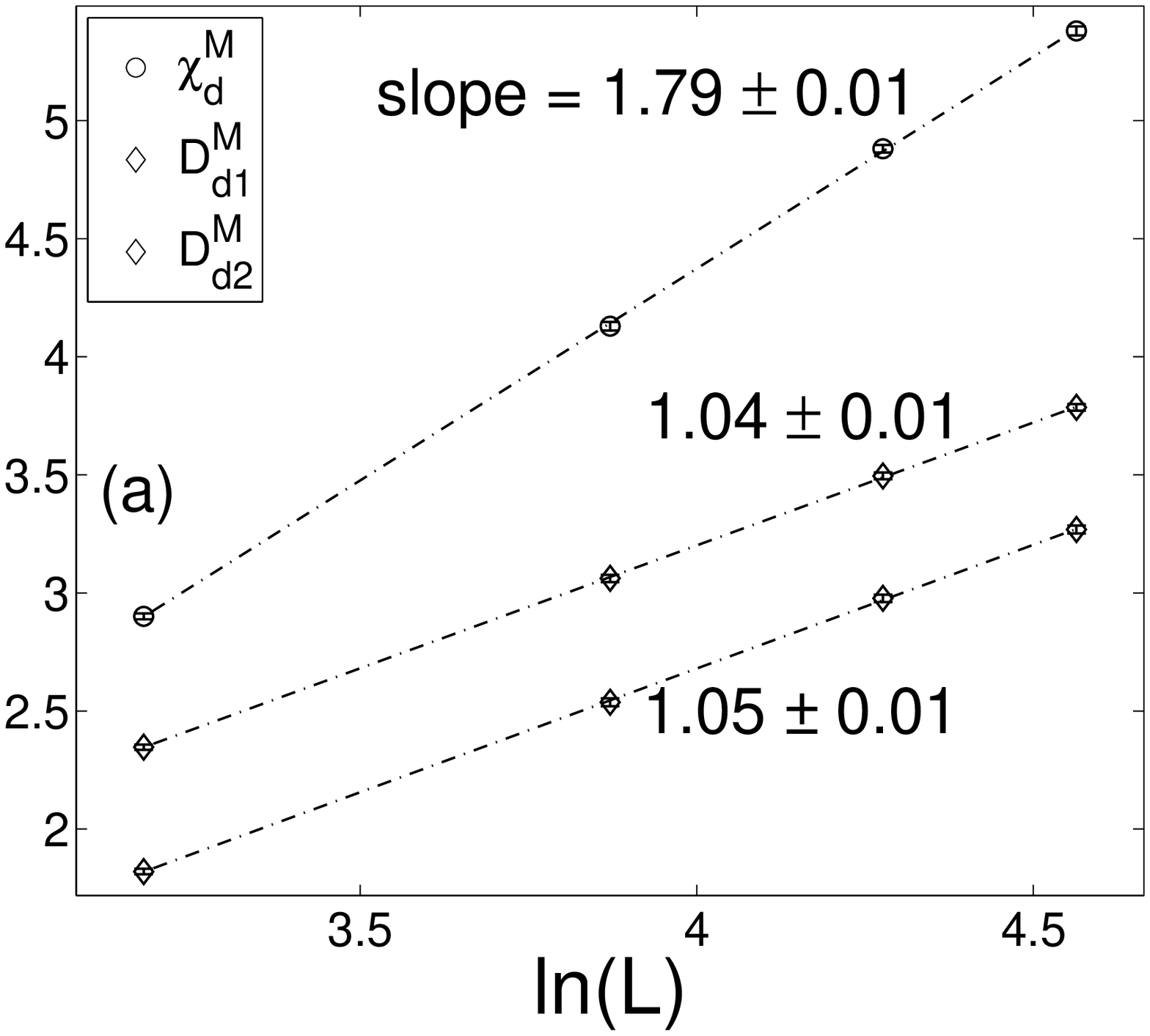}\label{fss_T0_75_P-D2}}
\subfigure{\includegraphics[scale=0.28,clip]{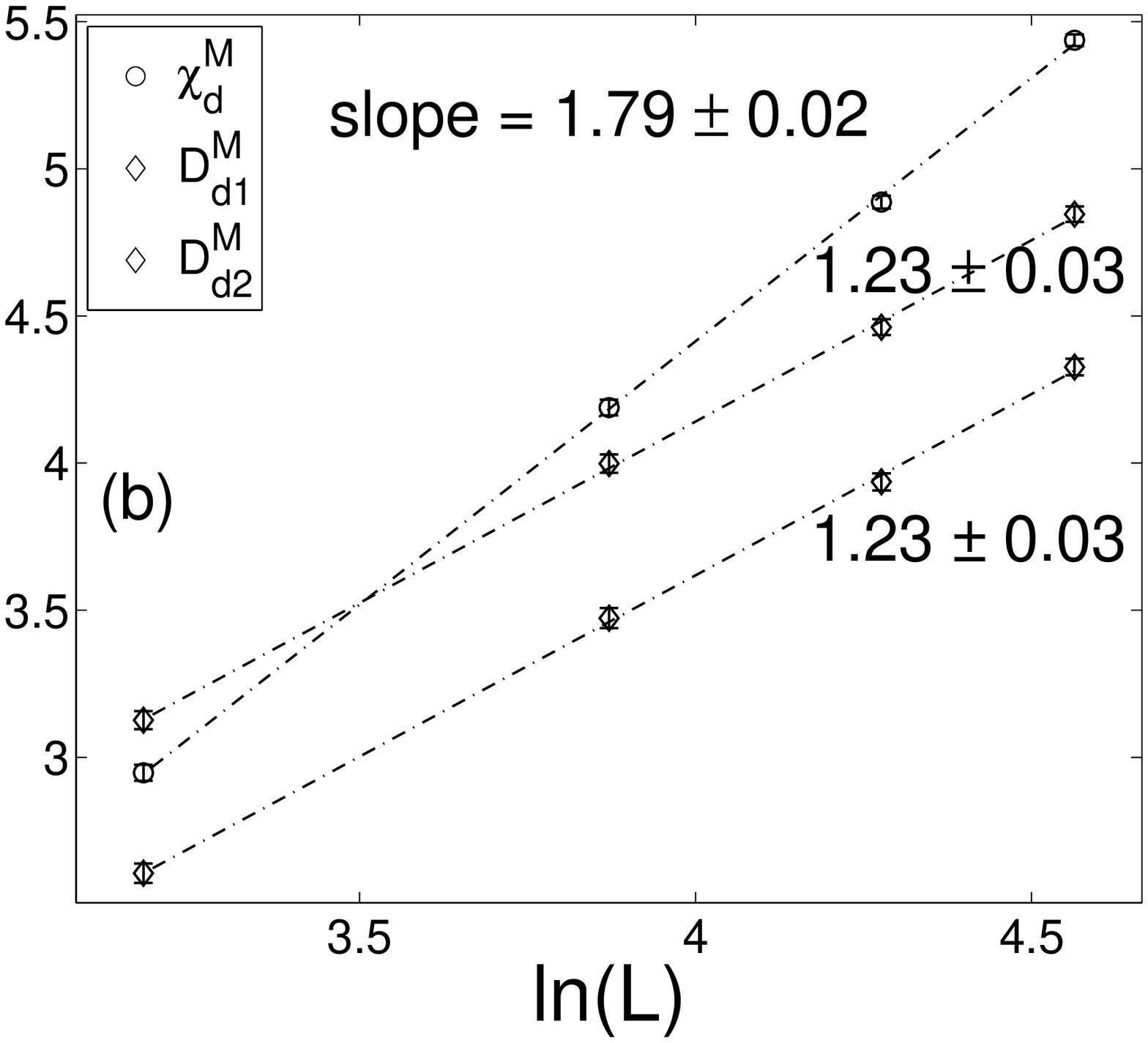}\label{fss_D2_P-D2_ma}}
\caption{The same as in Figure~\ref{fig:fss_uni} for the phase transitions near the $P$, $F_1$ and $FRM$ boundaries merging point: (a) $P \rightarrow FRM$ at $t=0.75$ and (b) $P \rightarrow FRM$ at $t=1$.} \label{fig:fss_nonuni} 
\end{figure}

\begin{figure}[t!]
\centering
\includegraphics[scale=0.5,clip]{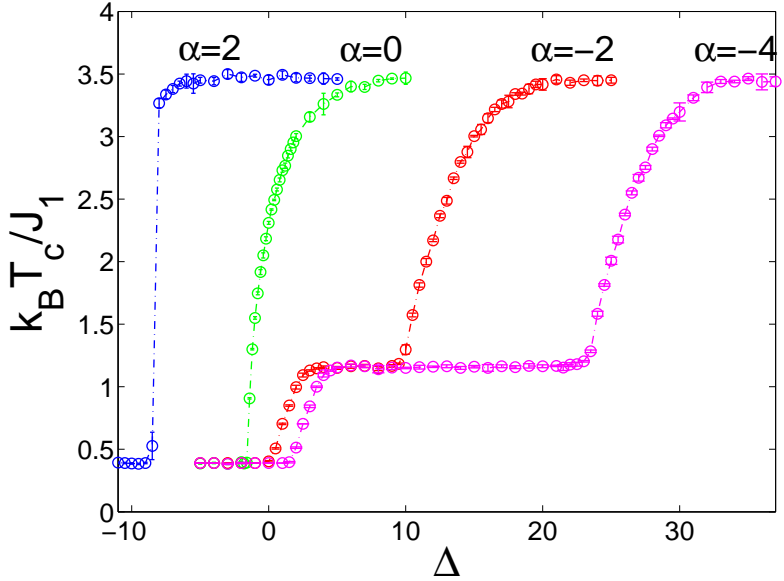}
\caption{Order-disorder phase diagrams for selected values of $\alpha$. The areas below and above the curves represent magnetically ordered and disordered phases, respectively.}\label{fig:PD}
\end{figure}

Based on the ground-state considerations above, let us first present the behavior of some relevant quantities in the parameters space where the identified phases are expected to appear. In particular, we choose the value of the biquadratic to bilinear exchange interaction ratio $\alpha=-2$ and investigate the thermodynamic quantities as functions of the temperature and the single-ion anisotropy.  
The former case is demonstrated in Fig.~\ref{fig:m_xi-T_J-2}, in which we show temperature dependencies of the direct dipolar order parameter (magnetization) $m_d$ and the corresponding susceptibility $\chi_d^M$ for selected values of the reduced single-ion anisotropy $\Delta$ and $L=48$. As expected from the minimum energy (\ref{Energy}) condition for $\alpha=-2$, the ground states are $F_1$ for $\Delta=0, 0.5$, $FRM$ for $\Delta=1, 9, 11$, and $F_2$ for $\Delta=12$, with the values of $m_d$ approaching $1/2, 1$ and $3/2$,
respectively, as $T \to 0$. As a result, for most values of $\Delta$ the curves show anomalies in the low-temperature region. Namely, thermal fluctuation can either markedly decrease (e.g., for $\Delta=1$) or even increase (e.g., for $\Delta=0.5$ or $11$) the magnetization. The respective magnetic orderings disappear at higher temperatures, which is manifested in the direct susceptibility peaks, presented in Fig.~\ref{fig:xi-T_J-2}. \\
\hspace*{5mm} Fig.~\ref{fig:x-D_J-2} demonstrates variations of the same quantities but now as functions of $\Delta$ for selected temperatures. In order to study quadrupolar ordering, we also include the behavior of the direct quadrupolar order parameter $q_d$ and the internal energy $e$ along with their respective response functions, the direct quadrupolar susceptibility $\chi_d^Q$ and the specific heat $C$. Thus we can see that, for example, for $t=1.5$ there is no magnetic ordering for $\Delta \lesssim 10$ but the ferriquadrupolar ordering\footnote{Sublattices A and B are predominantly populated with spins of the same magnitude but not sign, i.e., $|S_{\rm A}|=1/2$ and $|S_{\rm B}|=3/2$ or $|S_{\rm A}|=3/2$ and $|S_{\rm B}|=1/2$.} ($FRQ$) exists within $3 \lesssim \Delta \lesssim 10$. Moreover, transitions between different phases do not occur instantly but they seem to be spread within some $\Delta$ intervals. This is reflected in broader peaks of the response functions which beside a typical spike also feature a broader shoulder.\\
\hspace*{5mm} The phase diagram in $(t-\Delta)$ parameter space determined from the peak positions of the response functions pertaining to different order parameters is presented in Fig.~\ref{fig:PD_J2-2} for $\alpha=-2$. It features five different phases characterized by the following values of the order parameters: $P$ - paramagnetic with $m_{\rm A}=m_{\rm B}= 0$, $F_1$ - ferromagnetic with $m_{\rm A}=m_{\rm B}\neq 0\ (=1/2 \ {\rm at}\ T=0)$, $F_2$ - ferromagnetic with $m_{\rm A}=m_{\rm B}\neq 0\ (=3/2 \ {\rm at}\ T=0)$, $FRM$ - ferrimagnetic with $m_{\rm A} \neq m_{\rm B}\neq 0\ (= (1/2,3/2)$ or $(3/2,1/2) \ {\rm at}\ T=0)$, and $FRQ$ - ferriquadrupolar with $m_{\rm A}=m_{\rm B}=0$ and $q_{\rm A}\neq q_{\rm B}\neq 0$.\\
\hspace*{5mm} Next, we were interested in the character of the respective phase transitions between different phases. For this purpose we employed a FSS analysis, using the linear sizes $L=24,48,72$ and $96$. We selected several representative points on the phase boundaries and in Figs.~\ref{fig:F1-FRM_t015}-\ref{fig:FRQ-P-F2_t195} plotted $L$-dependent variations of some relevant quantities needed for FSS in the vicinity of those points. In the top rows we plot the order parameters relevant for the respective phases and the internal energy. We note that these quantities are little dependent on the lattice size and therefore only the curves for the largest size $L=96$ are presented. The lattice size dependence at criticality is best seen in the response functions shown in the bottom rows. In some cases, such as at the $F_1 \rightarrow FRM$ transition at $t=0.15$ in Fig.~\ref{fig:F1-FRM_t015}, the order parameter appears to change discontinuously and the corresponding staggered susceptibility shows a very narrow spike-like peak, which indicates possibility of a first-order phase transition. \\
\hspace*{5mm} However, in order to properly check the transition order we need to perform a FSS analysis employing the scaling relations (\ref{scalchi}-\ref{scalD2}). To obtain better quality data, we reran the simulations at the (pseudo)critical points estimated from the susceptibility peak locations, using up to $N=10^7$ MCS and employed the reweighting techniques~\cite{ferr88}. Such a way we could obtain various thermodynamic quantities used in the FSS analysis as continuous functions of model parameters, which allowed us a precise determination of the peaks maxima involved in the scaling relations (\ref{scalchi}-\ref{scalD2}). The log-log plots of these relations should give straight lines with the slopes corresponding to the respective critical exponents' ratios $\gamma/\nu$ and $1/\nu$ if the transition is second order. For the present model we expect the ratios consistent with the 2D Ising values $\gamma_I/\nu_I=7/4$ and $1/\nu_I=1$. On the other hand, in the case of a first-order transition the thermodynamic functions are expected to scale with volume, i.e., the slopes should be equal to $d=2$. Despite some visual first-order transition signatures, as described above, our analysis for the selected parameters did not confirm such a scenario and all the transitions were reliably evaluated as second order. Nevertheless, the critical exponents' ratios were not consistent with the Ising universality class in all the instances. In Fig.~\ref{fig:fss_uni} we show the cases in which the ratio $\gamma/\nu$ did not deviate from the Ising values beyond the error bars. However, in most of these cases the values of $1/\nu$ were not consistent with the standard value of 1 and varied with the model parameters. This finding would indicate that in fact these transitions comply with only weakly universal behavior~\cite{suzu74}. On the other hand, at the points near the merging of the phases $P$, $F_1$, $FRM$ and $FRQ$ (see the red dots in Fig.~\ref{fig:PD_J2-2}) both the exponent ratios $\gamma/\nu$ and $1/\nu$ deviate from the Ising values and thus violate universality. The log-log plots for these two cases are presented in Fig.~\ref{fig:fss_nonuni}. The values of $\gamma/\nu$ are larger than $\gamma_I/\nu_I$ beyond the error bars. We note that even clearer violation of Ising universality at the merging point of the $P$, $F_1$, $FRM$ and $FRQ$ phases was also observed in the tree-dimensional spin-3/2 BEG model on a simple cubic lattice simulated by the Creutz cellular automaton~\cite{sefe10}.\\
\hspace*{5mm} Finally, in order to verify the EFT predictions~\cite{kane93} about the discontinuous character of the order-disorder phase boundaries for positive and step-wise variation for larger negative values of the biquadratic to bilinear exchange interaction ratio $\alpha$, we ran simulations for several values of $\alpha$ and estimated the phase boundaries between the paramagnetic and ordered phases. The results for $\alpha=2,0,-2$ and $-4$ are presented in Fig.~\ref{fig:PD}. The discontinuous behavior for $\alpha=2$ is evident and thus in this case our MC simulations corroborate the EFT results. Nevertheless, except for the step associated with the $FRM$ phase there are no signs of any other steps for any value of $\alpha$. Therefore, the step-like dependence observed in the EFT calculations is likely just an artifact of the effective-field approximation.


\section{Conclusions}
\hspace*{5mm} In conclusion, we have studied the spin-3/2 Blume-Emery-Griffiths model on a honeycomb lattice by Monte Carlo simulations in order to verify some peculiar EFT predictions, as well as to investigate the character of the phase transitions between different phases. Our results confirmed discontinuous dependence of the order-disorder phase boundary as a function of a single-ion anisotropy for positive values of the exchange coupling ratio $\alpha$. However, the inexplicable multiple plateaus observed in the EFT calculations for negative $\alpha$ were not reproduced and thus we think they are merely artifacts of the used approximation.\\ 
\hspace*{5mm} Our finite-size scaling analysis, performed at several points of the phase diagram for a selected value of the biquadratic to bilinear exchange interaction ratio $\alpha=-2$, indicated that the phase transitions between different phases are of second order. However, the estimated values of the critical exponents' ratios pointed out to only weakly universal and in some points within the area where different boundaries merge even nonuniversal critical behavior. Similar universality violation was also recently observed in the three-dimensional BEG model by cellular automaton simulations~\cite{sefe10}.  

\section*{Acknowledgments}
This work was supported by the Scientific Grant Agency of Ministry of Education of Slovak Republic (Grant No. 1/0234/12). The authors acknowledge the financial support by the ERDF EU (European Union European Regional Development Fund) grant provided under the contract No. ITMS26220120047 (activity 3.2.).


\begin{thebibliography}{30}

\bibitem{blum71} M. Blume, V. Emery, R.B. Griffiths, Phys. Rev. A 4 (1971) 1071.
\bibitem{siva72} J. Sivardiere, M. Blume, Phys. Rev. B 5 (1972) 1126.
\bibitem{krin75} S. Krinsky, D. Mukamel, Phys. Rev. B 11 (1975) 399.
\bibitem{saba91} F.C. S\'{a} Barreto, O.F. de Alcantara Bonfim, Physica A 172 (1991) 378.
\bibitem{plas93} J.A. Plascak, J.G. Moreira, F.C. S\'{a} Barreto, Phys. Lett. A, 173 (1993), 360.
\bibitem{kane93} T. Kaneyoshi, M. Ja\v{s}\v{c}ur, Phys. Lett.A 177 (1993) 172.
\bibitem{bakk96} A. Bakkali, M. Kerouad, M. Saber, Physica A 229 (1996) 563.
\bibitem{peli96} L. Peliti, M. Saber, Phys. Stat. Sol. B 195 (1996) 537.
\bibitem{bakc93} A. Bakchich, A. Bassir, A. Benyoussef, Physica A 195 (1993) 188.
\bibitem{bakc01} A. Bakchich, M. El Bouziani, J. Phys. Condens. Matter 13 (2001) 91.
\bibitem{ilko96} V. Ilkovi\v{c}, Physica A, 234 (1996) 545.
\bibitem{tuck00} J.W. Tucker, J. Magn. Magn. Mater. 214 (2000) 121.
\bibitem{kesk08} M. Keskin, O. Canko, J. Magn. and Magn. Mater. 320 (2008) 8.
\bibitem{lara98} D. Pena Lara, J.A. Plascak, S.J. Ferreira, O.F. de Alcantara Bonfim, J. Magn. Magn. Mater., 177–81 (1998) 163.
\bibitem{sefe10} N. Sefero\v{g}lu, Commun. Comput. Phys. 7 (2010) 779.
\bibitem{wolf04} U. Wolff, Computer Physics Communications 156 (2004) 143.
\bibitem{blum66} M. Blume, Phys. Rev. 141 (1966) 517.
\bibitem{cape66} H. Capel, Physica (Amsterdam) 32 (1966) 966.
\bibitem{cape67} H. Capel, Physica (Amsterdam) 33 (1967) 295.
\bibitem{bekh97} S. Bekhechi, A. Benyoussef, Phys. Rev. B 56 (1997) 13954.
\bibitem{ferr88} A.M. Ferrenberg, R.H. Swendsen, Phys. Rev. Lett. 61 (1988) 2635.
\bibitem{suzu74} M. Suzuki, Progr. Theor. Phys. 51 (1974) 1992.

\end{thebibliography}
\end{document}